\RequirePackage{fix-cm}
\documentclass{svjour3}                     
\smartqed  
\usepackage{amsmath,epsfig}
\usepackage{latexsym}
\usepackage{amsfonts}
\usepackage{graphicx} 
\usepackage{subfigure}
\usepackage{amssymb}
\usepackage{stmaryrd}
\usepackage{caption}
\usepackage{color}

\newcommand{\dfn}{\triangleq}
\newcommand{\QED}{\Box} 
\newcommand{\rw}{\rightarrow} 
\newcommand{\Prob}{\mathbb{P}}    
\newcommand{\Real}{\mathbb{R}}  
  
\newcommand{\mB}{{\mathcal B}}
\newcommand{\mP}{{\mathcal P}}

\newcommand{\mC}{{\mathcal C}}

\newcommand{\mS}{{\mathcal S}}
\newcommand{\mI}{{\mathcal I}}

\newcommand{\mF}{{\mathcal F}}
\newcommand{\mE}{{\mathcal E}}
\newcommand{\mX}{{\mathcal X}}
\newcommand{\mN}{{\mathcal N}}

\newcommand{\mO}{{\mathcal O}}
\newcommand{\mT}{{\mathcal T}}

\newcommand{\mR}{{\mathcal R}}

\newcommand{\kS}{{\mathfrak S}}

\newcommand{\var}{\mbox{\sf Var}}

\newtheorem{Teorema}{\em Theorem}
\newtheorem{Definicion}{\em Definition}
\newtheorem{Lema}{\em Lemma}

\newtheorem{Corolario}{\em Corollary}
\newtheorem{Assumption}{\em Assumption}
\newtheorem{Nota}{\em Remark}
\newtheorem{Algoritmo}{\em Algorithm}

\begin{document}

\title{A simple scheme for the parallelisation of particle filters and its application to the tracking of complex stochastic systems}


\author{
	Dan Crisan \and
	Joaqu\'in M\'iguez \and
	Gonzalo~R\'ios
}


\institute{
	Dan Crisan \at Department of Mathematics, Imperial College London (UK).\\
	\email{d.crisan@imperial.ac.uk}           
	\and
	Joaqu\'in M\'iguez \at School of Mathematical Sciences, Queen Mary University of London (UK).\\
	\email{j.miguez@qmul.ac.uk}
	\and
	Gonzalo R\'ios \at Department of Signal Theory \& Communications, Universidad Carlos III de Madrid (Spain).\\
	\email{griosm@tsc.uc3m.es}
}

\date{Received: date / Accepted: date}

\maketitle

\begin{abstract}
Considerable effort has been devoted to the design of schemes for the parallel, or distributed, implementation of particle filters. The approaches vary from the totally heuristic to the mathematically well-principled. However, the former are largely based on (often loose) approximations that prevent the claim of any rigorous guarantees of convergence, whereas the latter involve considerable overhead to ensure the proper interaction of particles, which impairs the efficiency of the intended parallelisation. In this paper we investigate the use of possibly the simplest scheme for the parallelisation of the standard particle filter, that consists in splitting the computational budget into $M$ fully independent particle filters with $N$ particles each, and then obtaining the desired estimators by averaging over the $M$ independent outcomes of the filters. This approach minimises the parallelisation overhead yet displays highly desirable theoretical properties. Under very mild assumptions, we analyse the mean square error (MSE) of the estimators of 1-dimensional statistics of the optimal filtering distribution and show explicitly the effect of parallelisation scheme on the convergence rate. Specifically, we study the decomposition of the MSE into variance and bias components, to show that the former decays as $\frac{1}{MN}$, i.e., linearly with the total number of particles, while the latter converges towards $0$ as $\frac{1}{N^2}$. Parallelisation, therefore, has the obvious advantage of dividing the running times while preserving the (asymptotic) performance of the particle filter. Following this lead, we propose a time-error index to compare schemes with different degrees of parallelisation. Finally, we provide two numerical examples. The first one deals with the tracking of a Lorenz 63 chaotic system with dynamical noise and partial (noisy) observations, while the second example involves a dynamical network of modified FitzHugh-Nagumo (FH-N) stochastic nodes. The latter is a large dimensional system ($\approx3,000$ state variables in our computer experiments) designed to numerically reproduce typical electrical phenomena observed in the atria of the human heart. In both examples, we show how the proposed parallelisation scheme attains the same approximation accuracy as a centralised particle filter with only a small fraction of the running time, using a standard multicore computer.
\end{abstract}

\section{Introduction}

Over the past decade there has been a continued interest in the design of schemes for the implementation of particle filtering algorithms using parallel or distributed hardware of various types, including general purpose devices such as multi-core CPUs or graphical processing units (GPUs) \cite{Hendeby10}, and application-tailored devices such as field programmable gate arrays (FPGAs) \cite{Bolic05}. 

A particle filter is a recursive algorithm for the approximation of the sequence of posterior probability distributions that arise from a stochastic dynamical system in state-space form (see, e.g., \cite{Doucet01b,Ristic04,DelMoral04,Cappe07,Bain08,Kunsch13} and references therein for a general view of the field). A typical particle filter includes three steps that are repeated sequentially: 
\begin{itemize}
\item Monte Carlo sampling in the space of the state variables, 
\item computation of weights for the generated samples and, finally, 
\item resampling according to the weights. 
\end{itemize}
While~at~first~sight~the~algorithm~may look straightforward to parallelise (sampling and weighting can be carried out concurrently without any constraint), the resampling step involves the interaction of the whole set of Monte Carlo samples. Several authors have proposed schemes for `splitting' the resampling step into simpler tasks that can be carried out concurrently. The approaches are diverse and range from the heuristic \cite{Gelencser13,Miguez07,Hlinka12} to the mathematically well-principled \cite{Whiteley13,Verge13} (see also \cite{Bolic05} together with \cite{Miguez15,Heine15}). However, the former are largely based on (often loose) approximations that prevent the claim of any rigorous guarantees of convergence, whereas the latter involve non-negligible overhead to ensure the proper interaction of particles. 
The double bootstrap filter as described in \cite{Verge13}, for example, performs resampling at two levels (involving individual particles and sets of particles, respectively) and only one of the two admits direct parallelisation. The distributed resampling scheme of \cite{Bolic05,Miguez15,Heine15} is similar, as it resamples small subsets of particles in parallel, but the second level of resampling is substituted by an exchange of particles among different subsets which are typically assigned to different processing elements. The idea of exchanging particles is generalized in the $\alpha$ sequential Monte Carlo ($\alpha$-SMC) methodology \cite{Whiteley13}, where the resampling step is parameterised by designing a sequence of maps of interactions among particles. The higher-level resampling step of \cite{Verge13}, the particle exchange of \cite{Bolic05} or the parameterised interaction of \cite{Whiteley13} imply a computational overhead, i.e., there are extra computations that have to be performed in exchange for parallelising the original resampling task.

In this paper we investigate the use of possibly the simplest scheme for the parallelisation of the standard particle filter, that consists in splitting the computational budget into $M$ fully independent particle filters with $N$ particles each, and then obtaining the desired estimators by averaging over the $M$ independent outcomes of the filters. This approach minimises the parallelisation overhead, since there is no interaction at all among the filters, yet displays desirable theoretical properties. Under mild assumptions, we analyse the mean square error (MSE) of the estimators of 1-dimensional statistics of the optimal filtering distribution and show explicitly the effect of parallelisation scheme on the convergence rate. Specifically, we study the decomposition of the MSE into variance and bias components, to show that the former decays as $\frac{1}{MN}$, i.e., linearly with the total number of particles, while the latter converges towards $0$ as $\frac{1}{N^2}$. Independent parallelisation, therefore, has the obvious advantage of reducing the running time while preserving the (asymptotic) performance of a centralised filter. To compare different parallelisation schemes, we introduce a time-error index that brings together the time complexity (asymptotic order of the running time) and the estimation accuracy (asymptotic error rates) into a single quantitative figure of merit that can be used to compare schemes with different degrees of parallelisation. As a side result, we also show that the expected value of the random probability measure output by each independent filter converges in total variation distance to the true posterior with rate of order $1 / N$ (and we note that the average measure over the $M$ filters is just a sample-mean estimate of this expected measure).

The analysis of the particle filters in this paper is based on very mild assumptions on the stochastic dynamic model and classic induction arguments. We do not address uniform convergence over time or impose further assumptions on the models because our aim is to illuminate the relationship between the standard (centralised) particle filter and its parallelised versions in the simplest possible framework. A more sophisticated analysis can obviously be carried out by imposing additional assumptions on the dynamic models, yet the underlying argument for the comparison of parallel schemes would be identical. Additional results concerning the conditions that need to be satisfied to attain uniform convergence, the relationship with a $\alpha$ sequential Monte Carlo ($\alpha$-SMC) of \cite{Whiteley13} and a central limit theorem can be found in \cite{Heine15}. A study of ensembles of independent estimators in a machine learning framework, which shares the approach in this paper to some extent, is presented in \cite{Rosenblatt14}. While we have focused here on particle filters for discrete-time state-space models, the analysis can be similarly done for continuous-time systems and, indeed, the basic results needed for that case (the analysis of the approximation) can be found in \cite{Han13thesis}.  

The rest of the paper is organised as follows. In Section \ref{sBackground} we present basic background material and notations to be used through the rest of the paper. The theoretical results are introduced in Section \ref{sEnsembles}, including the analysis of the bias, MSE and total variation distance for various particle approximations. Numerical results for two examples are also presented and discussed, including the tracking of a stochastic Lorenz 63 system, in Section \ref{sLorenz}, and a stochastic-network model of an excitable medium, in Section \ref{sFHN}. The latter is a high-dimensional model (with $\sim$3,000 state variables in our simulations) that consists of a network of modified stochastic FitzHugh-Nagumo nodes, and displays some of the dynamical features observed during atrial fibrillation phenomena in the human heart \cite{Keener08a}. Finally, Section \ref{sConclusions} contains some concluding remarks.

\section{Background} \label{sBackground}

\subsection{Notation and preliminaries}

We first introduce some common notations to be used through the paper, broadly classified by topics. Below, $\Real$ denotes the real line, while for an integer $d\ge 1$, $\Real^d=\overbrace{\Real \times \ldots \times \Real}^{d \mbox{ {\tiny times}}}$

\begin{itemize}

\item Functions.
        \begin{itemize}
        \item The supremum norm of a real function $f:\Real^d \rw \Real$ is denoted as $\| f \|_\infty = \sup_{x\in\Real^d} | f(x) |$.
        \item $B(\Real^d)$ is the set of bounded real functions over $\Real^d$, i.e., $f \in B(\Real^d)$ if, and only if, $\| f \|_\infty < \infty$.
        \end{itemize}
        
\item Measures and integrals. Let $S \subseteq \Real^d$ be a subset of $\Real^d$.
        \begin{itemize}
        \item $\mB(S)$ is the $\sigma$-algebra of Borel subsets of $S$.
        \item $\mP(S)$ is the set of probability measures over the measurable space $(\mB(S),S)$.
        \item $(f,\mu) \dfn \int f(x) \mu(dx)$ is the integral of a real function $f:S \rw \Real$ with respect to (w.r.t.) a measure $\mu \in \mP(S)$.
        \item Given a probability measure $\mu \in \mP(S)$, a Borel set $A \in \mB(S)$ and the indicator function 
        $$
        I_A(x) = \left\{
                \begin{array}{ll}
                1, &\mbox{if } x \in A\\
                0, &\mbox{otherwise}
                \end{array}
        \right.,
        $$
        $\mu(A) = (I_A,\mu) = \int I_A(x) \mu(dx)$ 
        is the probability of $A$.

        \end{itemize}

        
\item Sequences, vectors and random variables (r.v.).
        \begin{itemize}
        \item We use a subscript notation for sequences, namely $x_{t_1:t_2} \dfn \{ x_{t_1}, \ldots, x_{t_2} \}$.
        \item For an element $x=(x_1,\ldots,x_d) \in \Real^d$ of an Euclidean space, its norm is denoted as $\| x \| = \sqrt{ x_1^2+\ldots+x_d^2 }$.
        \item The $L_p$ norm of a real r.v. $Z$, with $p \ge 1$, is written as $\| Z \|_p \dfn E[ |Z|^p ]^{1/p}$, where $E[\cdot]$ denotes expectation w.r.t. the distribution of $Z$.
        \end{itemize}
\end{itemize}


\subsection{State-space Markov models in discrete time} \label{ssModel}

Consider two random sequences, $\{ X_t \}_{t \ge 0}$ and $\{ Y_t \}_{t \ge 1}$, taking values in $\mX \subseteq \Real^{d_x}$ and $\Real^{d_y}$, respectively. Let $\Prob_t$ be the joint probability measure for the collection of random variables $\left\{ X_0, X_n, Y_n \right\}_{1 \le n \le t}$. 

We refer to the  sequence $\{ X_t \}_{t\ge 0}$ as the state (or signal) process and we assume that it is an inhomogeneous Markov chain governed by an initial probability measure $\tau_{0} \in \mP(\mX)$ and a sequence of Markov transition kernels $\tau_t : \mB(\mX) \times \mX \rw [0,1]$. To be specific, we define
\begin{eqnarray}
\tau_{0}(A) &\dfn& \Prob_0\left\{ X_0 \in A \right\}, \label{eqPrior} \\ 
\tau_t(A|x_{t-1}) &\dfn& \Prob_t\left\{ X_t \in A | X_{t-1}=x_{t-1} \right\}, \quad t \ge 1, \label{eqKernel}
\end{eqnarray}
where $A \in \mB(\mX)$ is a Borel set. The sequence $\{ Y_t \}_{t \ge 1}$ is termed the observation process. Each r.v. $Y_t$ is assumed to be conditionally independent of other observations given $X_t$, namely
\begin{eqnarray}
\Prob_t\left\{ Y_t \in A | X_{0:t} = x_{0:t}, \{ Y_k = y_k \}_{k \ne t} \right\} \\
 = \Prob_t\left\{ Y_t \in A | X_t = x_t  \right\} \nonumber
\end{eqnarray}
for any $A \in \mB(\Real^{d_y})$, and the conditional distribution of the r.v. $Y_t$ given $X_t=x_t$ is fully described by the probability density function (pdf) $g_t(y_t|x_t) > 0$. We often use $g_t$ as a function of $x_t$ (i.e., as a likelihood) and we emphasise this by writing $g_t^y(x) \dfn g_t(y|x)$. The prior $\tau_0$, the kernels $\{ \tau_t \}_{t \ge 1}$, and the functions $\{ g_t \}_{t \ge 1}$, describe a stochastic Markov state-space model in discrete time. 

The stochastic filtering problem consists in the computation of the posterior probability measure of the state $X_t$ given the sequence of observations up to time $t$. Specifically, for a given observation record $\{ y_t \}_{t \ge 1}$, we seek the probability measures
\begin{equation}
\pi_t(A) \dfn \Prob_t\left\{ 
	X_t \in A | Y_{1:t}=y_{1:t} 
\right\}, \quad t=0, 1, 2, ...
\nonumber
\end{equation}
where $A \in \mB(\mX)$. For many practical problems, the interest actually lies in the computation of statistics of $\pi_t$, e.g., the posterior mean or the posterior variance of $X_t$. Such statistics can be written as integrals of the form $(f,\pi_t)$, for some function $f:\mX\rw\Real$. Note that, for $t=0$, we recover the prior signal measure, i.e., $\pi_0=\tau_0$.

An associated problem is the computation of the one-step-ahead predictive measure 
\begin{equation}
\xi_t(A) \dfn \Prob_t\left\{ 
	X_t \in A | Y_{1:t-1}=y_{1:t-1} 
\right\}, \quad t = 1, 2, ...
\nonumber
\end{equation}
This measure can be explicitly written in terms of the kernel $\tau_t$ and the filter $\pi_{t-1}$. Indeed, for any integrable function $f:\mX\rw\Real$, we readily obtain (see, e.g., \cite[Chapter 10]{Bain08})
\begin{eqnarray}
(f,\xi_t) &=& \int \int f(x)\tau_t(dx|x')\pi_{t-1}(dx') \label{eqPredicting} \\ 
\nonumber &=&  \left(
	(f,\tau_t), \pi_{t-1}
\right),
\end{eqnarray}
and we write $\xi_t=\tau_t\pi_t$ as shorthand.

The filter at time $t$, $\pi_t$, can be obtained from the predictive measure, $\xi_t$, and the likelihood, $g_t^{y_t}$, by way of the so-called projective product \cite{Bain08}, or Boltzman-Gibbs transformation \cite{DelMoral04}, $\pi_t = g_t^{y_t} \star \xi_t$, defined as
\begin{equation}
(f,g_t^{y_t} \star \xi_t) \dfn \frac{
	(fg_t^{y_t},\xi_t)
}{
	(g_t^{y_t},\xi_t)
}
\nonumber
\end{equation}
for any integrable function  $f:\Real^{d_x}\rw\Real$, which, combined with \eqref{eqPredicting}, yields the recursive formula 
\begin{equation}
\pi_t = g_t^{y_t} \star \tau_t \pi_{t-1}.
\label{eqRecursiveFormula1}
\end{equation} 
It is also useful to keep track of the sequence of non-normalised measures $\{ \rho_t \}_{t \ge 0}$, where 
\begin{equation}
\rho_0 = \pi_0, \quad \rho_t = g_t^{y_t} \cdot \tau_t \rho_{t-1}
\label{eqDefRhoRec}
\end{equation}
and, for any integrable function $f:\mX\rw\Real$ and any measure $\alpha \in \mP(\mX)$, we define
\begin{equation}
(f,g_t^{y_t}\cdot \alpha) \dfn (fg_t^{y_t},\alpha).
\label{eqRecursiveFormula2}
\end{equation}
We remark that $\rho_t$ is {\em not} a probability measure, but an unnormalised version of $\pi_t$, namely
\begin{equation}
(f,\pi_t) = \frac{
	(f,\rho_t)
}{
	({\bf 1},\rho_t)
},
\nonumber
\end{equation}
where ${\bf 1}(x)=1$ is the constant unit function.

\subsection{Standard particle filter} \label{ssPFs}

Assume that a sequence of observations $Y_{1:T} = y_{1:T}$, for some $T<\infty$, is given. Then, the sequences of measures $\{ \pi_t \}_{t \ge 1}$, $\{ \xi_t \}_{t \ge 1}$ and $\{ \rho_t \}_{t \ge 0}$ can be numerically approximated using particle filtering. Particle filters are numerical methods based on the recursive relationships \eqref{eqRecursiveFormula1} and \eqref{eqRecursiveFormula2}. The simplest algorithm, often called `standard particle filter' or `bootstrap filter' \cite{Gordon93} (see also \cite{Doucet01}), can be described as follows.

\begin{Algoritmo} \label{alBF}
Bootstrap filter.
\begin{enumerate}
\item {\sf Initialisation.} At time $t=0$, draw $N$ i.i.d. samples,  $x_0^{(i)}$, $i=1,\ldots,N$, from the distribution $\tau_0$.
 
\item {\sf Recursive step.} Let $\{ x_{t-1}^{(i)} \}_{1 \le i \le N}$ be the particles (samples) generated at time $t-1$. At time $t$, proceed with the two steps below.
        \begin{enumerate}
        \item For $i=1,...,N$, draw a sample $\bar x_t^{(i)}$ from the probability distribution $\tau_t(\cdot|x_{t-1}^{(i)})$ and compute the normalised weight
        \begin{equation}
        w_t^{(i)} = \frac{
                g_t^{y_{t}}(\bar x_t^{(i)})
        }{
                \sum_{k=1}^N g_t^{y_{t}}(\bar x_t^{(k)})
        }.
        \end{equation}
        
        \item For $i=1,...,N$, let $x_t^{(i)}=\bar x_t^{(k)}$ with probability $w_t^{(k)}$, $k \in \{1,...,N\}$.
        \end{enumerate}
\end{enumerate}
\end{Algoritmo}

Step 2.(b) is referred to as resampling or selection. In the form stated here, it reduces to the so-called multinomial resampling algorithm \cite{Doucet00,Douc05} but the convergence of the filter can be easily proved for various other schemes (see, e.g., the treatment of the resampling step in \cite{Crisan01}). 

Using the sets $\{ \bar x_t^{(i)} \}_{1 \le i \le N}$ and $\{ x_t^{(i)} \}_{1 \le i \le N}$, we construct random approximations of $\xi_t$, $\rho_t$ and $\pi_t$, namely
\begin{eqnarray}
\xi_t^N &=& \frac{1}{N} \sum_{i=1}^N \delta_{\bar x_t^{(i)}}, \quad
\pi_t^N = \frac{1}{N} \sum_{i=1}^N \delta_{x_t^{(i)}}, \quad
\mbox{and} \label{eqDefs_measures1}\\ 
\rho_t^N &=& G_t^N \pi_t^N
\label{eqDefs_measures2}
\end{eqnarray}
where $\delta_{x}$ is the delta unit-measure located at $x \in \Real^{d_x}$ and\footnote{Note that $G^N_t$ is an estimate of the normalising constant for $\pi_t$ (namely, the integral $({\bf 1},\rho_t)$) which can be shown to be unbiased under mild assumptions \cite{DelMoral04}. In Bayesian model selection this constant is termed ``model evidence'' while in parameter estimation problems it is often referred to as the likelihood (of the unknown parameters) \cite{Andrieu10}.} 
\begin{equation}
G_t^N = \frac{1}{N^t} \prod_{k=1}^t \left(
	\sum_{j=1}^N g_k^{y_k}(\bar x_k^{(j)}) 
\right).
\label{eqGtN}
\end{equation}
For any integrable function $f$ on the state space, it is straightforward to approximate the integrals $(f,\xi_t)$, $(f,\pi_t)$ and $(f,\rho_t)$ as 
\begin{eqnarray}
(f,\xi_t) &\approx& (f,\xi_t^N) = \frac{1}{N} \sum_{i=1}^N f(\bar x_t^{(i)}), \nonumber\\ 
(f,\pi_t) &\approx& (f,\pi_t^N) = \frac{1}{N} \sum_{i=1}^N f(x_t^{(i)}), \quad \mbox{and} \nonumber \\
(f,\rho_t) &\approx& (f,\rho_t^N) = G_t^N (f,\pi_t^N),
\nonumber 
\end{eqnarray} 
respectively.

The convergence of particle filters has been analysed in a number of different ways. Here we use simple results for the convergence of the $L_p$ norms ($p\ge 1$) of the approximation errors. For the approximation of integrals w.r.t. $\xi_t$ and $\pi_t$ we have the following standard result. 

\begin{Lema} \label{lmConvBF}
Assume that the sequence of observations $Y_{1:T}=y_{1:T}$ is fixed (with $T<\infty$), $g_t^{y_{t}} \in B(\mX)$ and $g_t^{y_t}>0$ (in particular, $(g_t^{y_t},\xi_t) > 0$) for every $t=1,2,...,T$. Then for any $f \in B(\mX)$, any $p \ge 1$ and every $t = 1, \ldots, T$, 
\begin{eqnarray}
\left\|
        (f,\xi_t^N) - (f,\xi_t)
\right\|_p 
&\le&
        \frac{
                \bar c_t \| f \|_\infty
        }{
                \sqrt{N}        
        } \quad \mbox{and} \\ 
\left\|
        (f,\pi_t^N) - (f,\pi_t)
\right\|_p 
&\le& \frac{
	c_t \| f \|_\infty
}{
        \sqrt{N}        
},
\end{eqnarray}
where $\bar c_t$ and $c_t$ are finite constants independent of $N$, $\| f \|_\infty = \sup_{x \in \mX} |f(x)|<\infty$ and the expectations are taken over the distributions of the measure-valued random variables $\xi_t^N$ and $\pi_t^N$, respectively.
\end{Lema}

\noindent {\bf Proof:} This result is a special case of, e.g., Lemma 1 in \cite{Miguez13b}. $\QED$

\begin{Nota}
The constants $\bar c_t$ and $c_t$ can be easily shown to increase with $t$. It is possible to find error rates independent of $t$ by imposing additional assumptions on the state-space model (related to the stability of the optimal filter) \cite{DelMoral01c,DelMoral04}.
\end{Nota}

\section{Ensembles of independent particle filters} \label{sEnsembles}

%
\subsection{Overview}

Assume we run $M$ independent bootstrap filters (i.e., $M$ independent instances of Algorithm \ref{alBF}), with $N$ particles each, for the same sequence of observations $\{ y_t \}_{0 < t \le T}$. Each filter yields a random approximation $\pi_t^{i,N}$, $i=1, ..., M$, from which we compute the average $\pi_t^{M\times N} = \frac{1}{M}\sum_{i=1}^M \pi_t^{i,N}$ and adopt the mean square error (MSE) for bounded real test functions,
$
E\left[
	\left(
		(f,\pi_t^{M\times N}) - (f,\pi_t)
	\right)^2
\right], \quad f \in B(\mX),
$
as a performance metric. Since the underlying state-space model is the same for all filters and they are run in a completely independent manner, the measured-valued random variables $\pi_t^{i,N}$, $i=1, ..., M$, are i.i.d. and it is straightforward to show (via Lemma \ref{lmConvBF}) that
\begin{equation}
E\left[
	\left(
		(f,\pi_t^{M\times N}) - (f,\pi_t)
	\right)^2
\right] \le \frac{
	c_t^2 \| f \|_\infty^2
}{
	MN
}, \label{eqKK}
\end{equation}
for some constant $t$ independent of $N$ and $M$. However, the inequality \eqref{eqKK} falls short of characterising the performance of the parallelisation scheme because it does not illuminate the effect of the choice of $N$. In the extreme case of $N=1$, for example, $\pi_t^{M \times N}$ reduces to the outcome of a sequential importance sampling algorithm, with no resampling, which is known to degenerate quickly in practice. Instead of \eqref{eqKK}, we seek a bound for the approximation error that provides some indication on the trade-off between the number of independent filters, $M$, and the number of particles per filter, $N$.

For this purpose we derive bounds for the approximation error $\left(
	(f,\pi_t^{M \times N}) - (f,\pi_t)
\right)^2$ based on the classical decomposition of the MSE in variance and bias terms. First, we obtain preliminary results that are needed for the analysis of the average measure $\pi_t^{M \times N}$. In particular, we prove that the random unnormalised measure $\rho_t^N$ produced by the bootstrap filter (Algorithm \ref{alBF}) is unbiased and attains $L_p$ error rates proportional to $\frac{1}{\sqrt{N}}$, i.e., the same as $\xi_t^N$ and $\pi_t^N$. We use these results to derive an upper bound for the bias of $\pi_t^N$ which is proportional to $\frac{1}{N}$. The latter, in turn, enables us to deduce an upper bound for the MSE of the ensemble approximation $\pi_t^{M \times N}$ consisting of two additive terms that depend explicitly on $M$ and $N$. Specifically, we show that the variance component of the MSE decays linearly with the total number of particles, $K=MN$, while the bias term decreases quadratically with the number of particles per filter, i.e., with $N^2$.

All the results to be introduced in the rest of Section \ref{sEnsembles} hold true under the (mild) assumptions of Lemma \ref{lmConvBF}, which we summarise below for convenience of presentation.

\begin{Assumption} \label{asY}
The sequence of observations $Y_{1:T} = y_{1:T}$ is arbitrary but fixed, with $T<\infty$. 
\end{Assumption}

\begin{Assumption} \label{asG}
The likelihood functions are bounded and positive, i.e., 
$$
g_t^{y_t} \in B(\mX) \quad\mbox{and}\quad g_t^{y_t}>0 \quad\mbox{for every}\quad t= 1, 2, ..., T.
$$  
\end{Assumption}

\begin{Nota} \label{rmG}
Note that Assumptions \ref{asY} and \ref{asG} imply that 
\begin{itemize} 
\item $(g_t^{y_t},\alpha) > 0$, for any $\alpha \in \mP(\mX)$, and 
\item $\prod_{k=1}^T g_t^{y_t} \le \prod_{k=1}^T \| g_t^{y_t} \|_\infty < \infty$, 
\end{itemize}
for every $t=1, 2, ..., T$.
\end{Nota}

\begin{Nota} 
We seek simple convergence results for a fixed time horizon $T < \infty$, similar to Lemma \ref{lmConvBF}. Therefore, no further assumptions related to the stability of the optimal filter for the state-space model \cite{DelMoral01c,DelMoral04} are needed. If such assumptions are imposed then stronger (time uniform) asymptotic convergence rates can be found. See \cite{Heine15} for some additional results that apply to the independent filters $\pi_t^{i,N}$ and the ensemble $\pi_t^{M \times N}$.
\end{Nota}

%
\subsection{Bias and error rates} \label{ssBias&Rates}

Our analysis relies on some properties of the particle approximations of the non-normalised measures $\rho_t$, $t \ge 1$. We first show that the estimate $\rho_t^N$ in Eq. \eqref{eqDefs_measures1} and  \eqref{eqDefs_measures2} is unbiased.

\begin{Lema} \label{lmUnbiased}
If Assumptions \ref{asY} and \ref{asG} hold, then 
$$
E\left[
	(f,\rho_t^N)
\right] = (f,\rho_t)
$$
for any $f \in B(\mX)$ and every $t=1, 2, ..., T$.
\end{Lema}

\noindent {\bf Proof:} See Appendix \ref{apUnbiased}. $\QED$

Combining Lemma \ref{lmUnbiased} with the standard result of Lemma \ref{lmConvBF} leads to an explicit convergence rate for the $L_p$ norms of the approximation errors $(f,\rho_t^N) - (f,\rho_t)$. This is formally stated below.

\begin{Lema} \label{lmLpRho}
If Assumptions \ref{asY} and \ref{asG} hold, then, for any $f\in B(\mX)$, any $p\ge 1$ and every $t=1, 2, ..., T$, we have the inequality
\begin{equation}
\| (f,\rho_t^N) - (f,\rho_t) \|_p \le \frac{
	\tilde c_t \| f \|_\infty
}{
	\sqrt{N}
}, \label{eqStatementLpRho}
\end{equation}
where $\tilde c_t < \infty$ is a constant independent of $N$. 
\end{Lema}

\noindent {\bf Proof:} See Appendix \ref{apLpRho}. $\QED$

Finally, Lemmas \ref{lmUnbiased} and \ref{lmLpRho} together enable the calculation of explicit rates for the bias of the particle approximation of $(f,\pi_t)$. This is the first contribution of this paper and a key result for the decomposition of the MSE $E\left[ \left( (f,\pi_t^{M\times N}), (f,\pi_t) \right)^2\right]$ into variance and bias terms. To be specific, we can prove the following theorem.

\begin{Teorema}\label{thBias} 
If $0 < ({\bf 1},\rho_t) < \infty$ for $t=1, 2, ..., T$ and Assumptions \ref{asY} and \ref{asG} hold, then, for any $f\in B(\mX)$ and every $0 \le t \le T$, we obtain
$$
\left|
	E\left[
		(f,\pi_t^N) - (f,\pi_t)
	\right]
\right| \le \frac{
	\hat c_t \| f \|_\infty
}{
	N
},
$$
where $\hat c_t < \infty$ is a constant independent of $N$.
\end{Teorema}

\noindent {\bf Proof:} Let us first note that $(f,\pi_t) = (f,\rho_t)/({\bf 1},\rho_t)$ and
\begin{eqnarray}
(f,\pi_t^N) &=& \frac{
	(f,\rho_t^N)
}{
	G_t^N
} \label{eqSlip0} \\
&=& \frac{
	(f,\rho_t^N)
}{
	G_t^N ({\bf 1},\pi_t^N)
} \label{eqSlip1} \\
&=& \frac{
	(f,\rho_t^N)
}{
	({\bf 1},\rho_t^N)
}, \label{eqSlip2}
\end{eqnarray}
where \eqref{eqSlip0} follows from the construction of $\rho_t^N$, \eqref{eqSlip1} holds because $({\bf 1},\pi_t^N)=1$ and \eqref{eqSlip2} is, again, a consequence of the definition of $\rho_t^N$. Therefore, the difference $(f,\pi_t^N)-(f,\pi_t)$ can be written as
$$
(f,\pi_t^N)-(f,\pi_t) = \frac{
	(f,\rho_t^N)
}{
	({\bf 1},\rho_t^N)
} - \frac{
	(f,\rho_t)
}{
	({\bf 1},\rho_t)
}
$$
and, since $(f,\rho_t)=E[(f,\rho_t^N)]$ (from Lemma \ref{lmUnbiased}), the bias can be expressed as
\begin{equation}
E\left[
	(f,\pi_t^N)-(f,\pi_t) 
\right] = E\left[
	\frac{
		(f,\rho_t^N)
	}{
		({\bf 1},\rho_t^N)
	} - \frac{
		(f,\rho_t^N)
	}{
		({\bf 1},\rho_t)
	}
\right].
\label{eqSlip3}
\end{equation}
Some elementary manipulations on \eqref{eqSlip3} yield the equality
\begin{eqnarray}
E\hspace{-0.25em}\left[
	(f,\pi_t^N)-(f,\pi_t) 
\right] \hspace{-0.25em} = E\left[
	(f,\pi_t^N) \frac{
		({\bf 1},\rho_t) \hspace{-0.25em} - \hspace{-0.25em} ({\bf 1},\rho_t^N)
	}{
		({\bf 1},\rho_t)
	}
\right].
\label{eqSlip4}
\end{eqnarray}

If we realise that $E[ ({\bf 1},\rho_t) - ({\bf 1},\rho_t^N) ]=0$ (again, a consequence of Lemma \ref{lmUnbiased}) and move the factor $({\bf 1},\rho_t)^{-1}$ out of the expectation, then we easily rewrite Eq. \eqref{eqSlip4} as
\begin{eqnarray}
& E&\left[
	(f,\pi_t^N)-(f,\pi_t) 
\right] \nonumber\\
&=& \frac{1}{({\bf 1},\rho_t)} E\left[
	(f,\pi_t^N) \left(
		({\bf 1},\rho_t) - ({\bf 1},\rho_t^N)
	\right)
\right] \nonumber \\
& & - \frac{
	(f,\pi_t)
}{
	({\bf 1},\rho_t)
} E\left[ 
	({\bf 1},\rho_t) - ({\bf 1},\rho_t^N) 
\right] \nonumber\\
&=& \frac{1}{({\bf 1},\rho_t)} E\left[
	\left(
		(f,\pi_t^N)-(f,\pi_t) 
	\right)\left(
		({\bf 1},\rho_t)-({\bf 1},\rho_t^N) 
	\right)
\right] \nonumber \\
&\le& \frac{1}{({\bf 1},\rho_t)} \sqrt{
	E\left[
		\left(
			(f,\pi_t^N)-(f,\pi_t) 
		\right)^2
	\right] 
	} \nonumber \\
& & \times	\sqrt{
	E\left[
		\left(
			({\bf 1},\rho_t)-({\bf 1},\rho_t^N) 
		\right)^2
	\right]
} \label{eqSlip5} \\
&\le& \frac{1}{({\bf 1},\rho_t)} \left(
	\frac{
		c_t \| f \|_\infty
	}{
		N
	} \times \frac{
		\tilde c_t
	}{
		N
	}
\right) = \frac{
	\hat c_t \| f \|_\infty
}{
	N
}, \label{eqSlip6}
\end{eqnarray}

where we have applied the Cauchy-Schwartz inequality to obtain \eqref{eqSlip5}, \eqref{eqSlip6} follows from Lemmas \ref{lmConvBF} and \ref{lmLpRho} and the constant 
$$
\hat c_t =  \frac{
	c_t \tilde c_t \| f \|_\infty
}{
	({\bf 1},\rho_t)
} < \infty
$$
is independent of $N$. $\QED$

For any $f \in B(\mX)$, let $\mE_t^N(f)$ denote the approximation difference, i.e.,
$$
\mE_t^N(f) \dfn (f,\pi_t^N) - (f,\pi_t).
$$
This is a random variable whose second order moment yields the MSE of $(f,\pi_t^N)$. It is straightforward to obtain a bound for the MSE from Lemma \ref{lmConvBF} and, by subsequently using Theorem \ref{thBias}, one also readily finds a similar bound for the variance of $\mE_t^N(f)$, denoted $\var[\mE_t^N(f)]$. These results are explicitly stated by the corollary below.

\begin{Corolario} \label{coVar}
If $0<({\bf 1},\rho_t)<\infty$ for $t=1, 2, ..., T$ and Assumptions \ref{asY} and \ref{asG} hold, then, for any $f\in B(\mX)$ and any $0 \le t \le T$, we obtain
\begin{eqnarray}
E\left[
	\left(
		\mE_t^N(f)
	\right)^2
\right] &\le& \frac{
	c_t^2 \| f \|_\infty^2
}{
	N
} \quad \mbox{and} \label{eqTrivial}\\
\var\left[
	\mE_t^N(f)
\right] &\le& \frac{
	\left( c_t^v \right)^2 \| f \|_\infty^2 
}{
	N
},\label{eqBoundVar}
\end{eqnarray}
where $c_t$ and $c_t^v$ are finite constants independent of $N$.
\end{Corolario}

\noindent {\bf Proof:} The inequality \eqref{eqTrivial} for the MSE is a straightforward consequence of Lemma \ref{lmConvBF}. Moreover, we can write the MSE in terms of the variance and the square of the bias, which yields
\begin{eqnarray}
E\left[
	\left(
		\mE_t^N(f)
	\right)^2 	
\right] &=& \var\left[
	\mE_t^N(f)
\right] + E^2\left[
	\mE_t^N
\right] \nonumber \\
 &\le& \frac{
	c_t^2 \| f \|_\infty^2
}{
	N
}.
\label{eqVarBias}
\end{eqnarray}
Since Theorem \ref{thBias} ensures that $| E[\mE_t^N] | \le \frac{\hat c_t\| f \|_\infty}{N}$, then the inequality \eqref{eqVarBias} implies that there exists a constant $c_t^v<\infty$ such that \eqref{eqBoundVar} holds. $\QED$

%
\subsection{Error rate for the ensemble approximation} \label{ssEnsemble}

Let us run $M$ independent particle filters with the same (fixed) sequence of observations $Y_{1:T} = y_{1:T}$, $T<\infty$, and $N$ particles each. The random measures output by the $m$-th filter are denoted $\xi_t^{m,N}$, $\pi_t^{m,N}$ and $\rho_t^{m,N}$, with $m = 1, 2, ..., M$. Obviously, all the theoretical properties established in Section \ref{ssBias&Rates}, as well as the basic Lemma \ref{lmConvBF}, hold for each one of the $M$ independent filters.

\begin{Definicion} \label{defEnsemble}
The ensemble approximation of $\pi_t$ with $M$ independent filters is the discrete random measure $\pi_t^{M \times N}$ constructed as the average 
$$
\pi_t^{M \times N} = \frac{1}{M} \sum_{m=1}^M \pi_t^{m,N}.
$$
\end{Definicion}

It is apparent that similar ensemble approximations can be given for $\xi_t$ and $\rho_t$. Moreover, the statistical independence of the particle filters yields the following corollary as a straightforward consequence of Theorem \ref{thBias} and Corollary \ref{coVar}.

\begin{Corolario} \label{coMSE}
If $0 < ({\bf 1},\rho_t)<\infty$ for $t=1, 2, ..., T$ and Assumptions \ref{asY} and \ref{asG} hold, then, for any $f\in B(\mX)$ and any $0 \le t \le T$, the inequality
\begin{equation}
E\hspace{-0.25em}\left[
	\left(
		(f,\pi_t^{M \times N}) \hspace{-0.1em} - \hspace{-0.1em} (f,\pi_t)
	\right)^2
\right] \hspace{-0.05em} \le \hspace{-0.05em} \frac{
	(c_t^v)^2 \| f \|_\infty^2
}{
	MN
} \hspace{-0.05em} + \hspace{-0.05em} \frac{
	\hat c_t^2 \| f \|_\infty^2
}{
	N^2
} \label{eqCor2}
\end{equation}
holds for some constants $c_t^v$ and $\hat c_t$ independent of $N$ and $M$.
\end{Corolario}

\noindent {\bf Proof:} Let us denote 
\begin{eqnarray}
\mE_t^{M \times N}(f) &=& (f,\pi_t^{M \times N}) - (f,\pi_t) 
\quad \mbox{and} \nonumber \\ 
\mE_t^{m,N}(f) &=& (f,\pi_t^{m, N}) - (f,\pi_t)
\nonumber
\end{eqnarray}
for $m=1, 2, ..., M$. Since $\pi_t^{M \times N}$ is a linear combination of i.i.d. random measures, we easily obtain that
\begin{eqnarray}
\left|
	E\left[ 
		\mE_t^{M \times N}(f) 
	\right] 
\right|^2 &=& \left|
	\frac{1}{M} \sum_{m=1}^M E\left[
		\mE_t^{m,N}(f)
	\right]
\right|^2 \nonumber \\
&=& \left|
	E\left[ 
		\mE_t^{m,N}(f) 
	\right]
\right|^2 \nonumber \\ 
&\le& \frac{\hat c_t \| f \|_\infty}{N}, \quad \mbox{for any $m \le M$},
\label{eqP2}
\end{eqnarray}
where the inequality follows from Theorem \ref{thBias}. Moreover, again because of the independence of the random measures, we readily calculate a bound for the variance of $\mE_t^{M \times N}(f)$, 
\begin{equation}
\var\left[
	\mE_t^{M \times N}(f)
\right] = \frac{1}{M} \var\left[
	\mE_t^{m,N}(f)
\right] \le \frac{
	(c_t^v)^2 \| f \|_\infty^2
}{
	MN
},
\label{eqP1}
\end{equation}
where the inequality follows from Corollary \ref{coVar}. Since $E[ (\mE_t^{M \times N})^2 ] = \var[ \mE_t^{M\times N} ] + \left| E[ \mE_t^{M \times N} ] \right|^2$, combining \eqref{eqP1} and \eqref{eqP2} yields \eqref{eqCor2} and concludes the proof. $\QED$

The inequality in Corollary \ref{coMSE} is the main theoretical result in this paper and it admits several interpretations. Assume that some ``computational budget'' is given, i.e., that we have the resources to generate and update over time a fixed number $K$ of particles. If we run Algorithm \ref{alBF} with $K$ particles, then, according to Lemma \ref{lmConvBF}, the MSE $E[(\mE_t^K)^2]$ vanishes asymptotically with rate $\mO(1/K)$. The same result is obtained via Corollary \ref{coMSE}: a centralised particle filter with $K$ particles is the degenerate case of the ensemble approximation in Definition \ref{defEnsemble} with $M=1$ and $N=K$, hence Corollary \ref{coMSE} also states that the MSE decreases with rate $\mO(1/K)$. However, Corollary \ref{coMSE} also yields the same MSE rate when a proper ensemble is constructed, with some $M>1$ and $N=K/M$, which is the property that turns out relevant for parallelisation.

\begin{Nota} \label{rmCPFvsIPF}
Some key features of the proposed scheme become apparent from Corollary \ref{coMSE}:
	\begin{enumerate}
	\item If the total number of particles $K$ is fixed, then the ensemble approximation $\pi_t^{M \times N}$ with $K=MN$ and the standard (bootstrap filter) approximation with $K$ particles, $\pi_t^K$, yield the same MSE rate $\mO(1/K)$. Since the ensemble approximation is obtained by averaging over $M$ independent particle filters, the interactions among particles are strictly constrained to be local (within the $M$ subsets of size $N$ assigned to the independent particle filters). This implies that if the $M$ particle filters are run in parallel, there is no parallelisation overhead in terms of interaction among the parallel processing units.
	
	\item According to the inequality \eqref{eqCor2}, the bias of the estimator $(f,\pi_t^{M\times N})$ is controlled by the number of particles per subset, $N$, and converges quadratically, while, for fixed $N$, the variance decays linearly with $M$. The MSE rate is $\mathcal{O}\left( \frac{1}{MN} \right)$ as long as $N \ge M$. Otherwise, the term $\frac{\hat c_t^2 \| f \|_\infty^2}{N^2}$ becomes dominant and the resulting asymptotic error bound turns out higher.
	\end{enumerate}
\end{Nota}

\begin{Nota}
While the convergence results presented here have been proved for the standard bootstrap filter, it is straightforward to extend them to other classes of particle filters for which Lemmas \ref{lmConvBF} and \ref{lmUnbiased} hold. This includes most standard algorithms, including the auxiliary particle filter \cite{Pitt01} for which numerical results are reported in Section \ref{sFHN}.
\end{Nota}

%
\subsection{Comparison of parallelisation schemes via time--error indices} \label{ssTimeError}

The advantage of parallel computation is the drastic reduction of the time needed to run the particle filter. Let the running time for a particle filter with $K$ particles be of order $\mT(K)$, where $\mT:\mathbb{N}\rw(0,\infty)$ is some strictly increasing function of $K$. The quantity $\mT(K)$ includes the time needed to generate new particles, weight them and perform resampling. The latter step is the bottleneck for parallelisation, as it requires the interaction of all $K$ particles. Also, a ``straightforward'' implementation of the resampling step leads to an execution time $\mT(K)=K\log(K)$, although efficient algorithms exist that achieve to a linear time complexity, $\mT(K)=K$. We can combine the MSE rate 
and the time complexity 
to propose a a time--error performance metric. 

\begin{Definicion} \label{defMetric}
We define the time--error index of a particle filtering algorithm with running time of order $\mT$ and asymptotic MSE rate $\mR$ as
$
\mC \dfn \mT \times \mR.
$
\end{Definicion}

The smaller the index $\mC$ for an algorithm, the more (asymptotically) efficient its implementation. For the standard (centralised) particle filter with $K$ particles, the running time is of order $\mT(K)=K$ and the MSE rate is of order $\mR(K)=K^{-1}$, hence the time--error index becomes
$$
\mC_{spf}(K) = \mT(K) \times \mR(K) = 1.
$$
For the computation of the ensemble approximation $\pi_t^{M \times N}$ we can run $M$ independent particle filters in parallel, with $N=K/M$ particles each and no interaction among them. Hence, the execution time becomes of order $\mT(M,N)=N$. Since the error rate for the ensemble approximation is of order $\mR(M,N)=\left(\frac{1}{MN}+\frac{1}{N^2}\right)$, the time--error index of the ensemble approximation is
$$
\mC_{ipf}(M,N) = \mT(M,N) \times \mR(M,n) = \frac{1}{M} + \frac{1}{N}
$$
and hence it vanishes with $M, N \rw \infty$. In particular, since we have to choose $N \ge M$ to ensure a rate of order $\frac{1}{MN}$, then $\lim_{M \rw \infty} \mC_{ipf} = 0$. In any case, whenever $N>1$ it is apparent that $\mC_{ipf} < \mC_{spf}$. Similar comparisons can be carried out for other parallel particle filtering schemes as long as it is possible to identify the effect of the overhead in the running time $\mT(M,N)$.

\subsection{Expectation of the approximate filter}

Besides the computational view of Sections \ref{ssEnsemble} and \ref{ssTimeError}, Theorem \ref{thBias} can also be exploited to assess the ensemble approximation $\pi_t^{M \times N}$ in terms of the total variation distance defined as follows.

\begin{Definicion} \label{defTVD}
Let $\alpha, \beta \in \mP(\mX)$ be two probability measures. The total variation distance (TVD) between $\alpha$ and $\beta$ is usually defined as
$$
d_{TV}(\alpha,\beta) \dfn \sup_{ A \in \mB(\mX) } | (I_A,\alpha)-(I_A,\beta) |,
$$
where $I_A$ is the indicator function
$$
I_A(x) = \left\{
	\begin{array}{cl}
	1, &\mbox{if $x\in A$},\\
	0, &\mbox{otherwise.}
	\end{array}
\right.
$$
\end{Definicion}

If we regard the measures $\pi_t^{m,N}$ as i.i.d. realisations of the measure-valued r.v. $\pi_t^N$ (which is the random outcome of a bootstrap filter with $N$ particles), then the ensemble $\pi_t^{M \times N} = \frac{1}{M}\sum_{m=1}^M \pi_t^{m,N}$ can be interpreted as a sample mean approximation of the expectation of $\pi_t^N$. Indeed, if we introduce
\begin{equation}
\hat \pi_t^N \dfn E[ \pi_t^N ] = E[ \pi_t^{m,N} ] \quad \mbox{for every $m$},
\nonumber
\end{equation}
then it is apparent that, for any $f \in B(\mX)$, 
\begin{equation}
E\left[ (f,\pi_t^N) - (f,\pi_t) \right] = (f,\hat \pi_t^N) - (f,\pi_t),
\nonumber
\end{equation}
and $(f,\pi_t^{M \times N}) = \frac{1}{M}\sum_{m=1}^M (f,\pi_t^{m,N})$ is the sample-mean approximation of $(f,\hat \pi_t^N)$. Theorem \ref{thBias} can be re-stated immediately as follows.

\begin{Corolario}\label{coTVD} 
If $0 < ({\bf 1},\rho_t) < \infty$ for $t=1, 2, ..., T$ and Assumptions \ref{asY} and \ref{asG} hold, then, for every $0 \le t \le T$, we obtain that
$$
d_{TV}\left(
	\hat \pi_t^N, \pi_t
\right) \le \frac{
	\hat c_t
}{
	N
},
$$
where $\hat c_t < \infty$ is a constant independent of $N$.
\end{Corolario}

\noindent {\bf Proof:} From the definition of total variation distance,
\begin{eqnarray}
d_{TV}\left(
	\hat \pi_t^N, \pi_t
\right) &=& \sup_{ A \in \mB(\mX) } \left|
	(I_A,\hat \pi_t^N) - (I_A,\pi_t)
\right| \nonumber \\ 
&\le& 
\sup_{ f \in B(\mX) : \| f \|_\infty \le 1 } \left|
	(f,\hat \pi_t^N) - (f,\pi_t)
\right|, \label{eqTVD0}
\end{eqnarray}
since $I_A \in B(\mX)$ and $\| I_A \|_\infty = 1$ for every Borel set $A$. However, $(f,\hat \pi_t^N) - (f,\pi_t) = E\left[ (f,\pi_t^N) - (f,\pi_t) \right]$, hence a straightforward application of Theorem \ref{thBias} completes the proof.  $\QED$

%


\section{Example: Stochastic Lorenz 63 model} \label{sLorenz}

\subsection{The 3-dimensional Lorenz system} \label{ssLorenz}

Let us illustrate the numerical performance of the proposed independent parallelisation scheme by means of some computer simulations. First, we consider the problem of tracking the state of a 3-dimensional Lorenz system \cite{Lorenz63} with additive dynamical noise and partial observations \cite{Chorin04}. To be specific, consider a 3-dimensional stochastic process $\{ X(s) \}_{s\in(0,\infty)}$ ($s$ denotes continuous time) taking values on $\Real^3$, whose dynamics is described by the system of stochastic differential equations
\begin{eqnarray}
dX_1 &=& -{\sf s} (X_1-Y_1) + dW_1, \nonumber\\
dX_2 &=& {\sf r} X_1 - X_2 - X_1X_3 + dW_2, \nonumber\\
dX_3 &=& X_1X_2 - {\sf b} X_3 + dW_3, \nonumber
\end{eqnarray}
where $\{ W_i(s) \}_{s\in(0,\infty)}$, $i=1, 2, 3$, are independent 1-dimensional Wiener processes and 
$$
({\sf s,r,b}) = \left(
	10, 28, \frac{8}{3}
\right)
$$ 
are static model parameters\footnote{Note the difference in notation between the continuous time $s$ and the parameter $\sf s$.} (which yield chaotic dynamics). A discrete-time version of the latter system using Euler's method with integration step $T_d = 10^{-3}$ is straightforward to obtain and yields the model
\begin{eqnarray}
X_{1,n} &=& X_{1,n-1} - T_d {\sf s} (X_{1,n-1}-X_{2,n-1}) \nonumber \\
&&+ \sqrt{T_d} U_{1,n},\label{eqDiscreteLorenz-1}\\
X_{2,n} &=& X_{2,n-1} + T_d ( {\sf r} X_{1,n-1} - X_{2,n-1} - X_{1,n-1}X_{3,n-1} ) \nonumber \\
&&+ \sqrt{T_d} U_{2,n}, \label{eqDiscreteLorenz-2}\\
X_{3,n} &=& X_{3,n-1} + T_d ( X_{1,n-1}X_{2,n-1} - {\sf b} X_{3,n-1} ) \nonumber \\
&&+ \sqrt{T_d} U_{3,n}, \label{eqDiscreteLorenz-3}
\end{eqnarray} 
where $\{ U_{i,n} \}_{n=0, 1, ...}$, $i=1,2,3$, are independent sequences of i.i.d. normal random variables with 0 mean and variance 1. System \eqref{eqDiscreteLorenz-1}-\eqref{eqDiscreteLorenz-3} is partially observed every 100 discrete-time steps. Specifically, we collect a sequence of scalar observations $\{ Y_t \}_{t=1, 2, ...}$, of the form 
\begin{equation}
Y_t = X_{1,100t} + V_t,  
\label{eqObservLorenz}
\end{equation}
where $\{ V_t \}_{t=1, 2, ...}$ is a sequence of i.i.d. normal random variables with zero mean and variance $\sigma^2 = \frac{1}{2}$.

Let $X_n=(X_{1,n},X_{2,n},X_{3,n}) \in \Real^3$ be the state vector at discrete time $n$. The dynamic model given by Eqs. \eqref{eqDiscreteLorenz-1}--\eqref{eqDiscreteLorenz-3} yields the family of kernels $\tau_{n,\theta}(dx|x_{n-1})$ and the observation model of Eq. \eqref{eqObservLorenz} yields the likelihood function 
$$
g_{t,\theta}^{y_t}(x_{100t}) \propto \exp\left\{ 
	-\frac{1}{2\sigma^2}\left(
		y_t - x_{1,100t}
	\right)^2
\right\},
$$
both in a straightforward manner. The goal is to track the sequence of joint posterior probability measures $\pi_t$, $t=1, 2, ...$, for $\{ \hat X_t \}_{t=1, ...}$, where $\hat X_t = X_{100t}$. Note that one can draw a sample $\hat X_t = \hat x_t$ conditional on $\hat X_{t-1} = \hat x_{t-1}$ by successively simulating
$$
\tilde x_n \sim \tau_{n,\theta}(dx|\tilde x_{n-1}), \quad n=100(t-1)+1, ..., 100t, 
$$
where $\tilde x_{100(t-1)} = \hat x_{t-1}$ and $\hat x_t = \tilde x_{100t}$. The prior measure for the state variables is normal, namely
$$
X_0 \sim \mN(x_*,v_0^2 \mI_3),
$$
where $x_* = (-10.2410; -1.3984; -23.6752)$ is the mean\footnote{Chosen from a typical trajectory of the {\em deterministic} Lorenz 63 model.} and $v_0^2\mI_3$ is the covariance matrix, with $v_0^2 = 10$ and $\mI_3$ the 3-dimensional identity matrix.

\subsection{Simulation setup} \label{ssSetup}


We aim at illustrating the gain in relative performance, taking into account both estimation errors and running time, that can be attained using ensembles of independent particle filters. With this purpose, we have applied 
\begin{itemize}
\item the standard bootstrap filter (Algorithm \ref{alBF}), termed BF in the sequel,
\item the double bootstrap method as described in \cite{Verge13}, and
\item the ensemble of independent bootstrap filters (BFs) that we have investigated in Section \ref{sEnsembles}
\end{itemize}
to track the sequence of probability measures $\pi_t$ generated by the 3-dimensional Lorenz model described in Section \ref{ssLorenz}. We have generated a sequence of $200$ synthetic observations, $\{ y_t; t=1, ..., 200 \}$, spread over an interval of 20 continuous time units, corresponding to $2 \times 10^4$ discrete time steps in the Euler scheme (hence, one observation every 100 steps). The same observation sequence has been used for all the simulations. 

The ensemble of independent particle filters consists of $M$ filters with $N$ particles each, abiding by the notation in Section \ref{sEnsembles}, with resampling for every $t=1, 2, ...$, i.e., every time an observation is collected and processed to obtain importance weights. Since the time scale of the Euler approximation of Eqs. \eqref{eqDiscreteLorenz-1}--\eqref{eqDiscreteLorenz-3} is $n=100t$, a resampling step is taken every 100 steps of the underlying discrete-time system. Similarly, the double bootstrap algorithm consists of $M$ subsets of particles ({\em particle islands}, in the terminology of \cite{Verge13}) with $N$ particles per subset. Within each subset, the $N$ particles are resampled for every $t=1, 2, ...$, while the subsets are resampled for $t = 5k$, $k=1, 2, ...$. The local resampling (of the particles in the same subset) can be carried out in parallel for the $M$ islands. The standard bootstrap filter runs with $K$ particles, where $K=MN$ (unless explicitly stated) for a fair comparison, and resampling for $t=1, 2, ...$.

We have coded the three algorithms in Matlab (version 7.11.0.584 [R2010b] with the parallel computing toolbox) and run the experiments using a pool of identical multi-processor machines, each one having 8 cores at 3.16 GHz and 32 GB of RAM memory. The standard (centralised) BF is run with $K=NM$ particles in a single core. For the ensemble of independent particle filters we allow the parallel computing toolbox to allocate all available cores per server in order to run all BFs concurrently. For the double BF method, we allow to run the separate BFs in parallel up to the subset-level resampling steps. 

To assess the approximation errors, we have computed empirical MSEs for the approximation of the posterior mean, $E[\hat X_t | Y_{1:t}] = (I,\pi_t)$, where $I(x)=x$ is the identity function, for the three algorithms at the last update step, $t=200$. Note, however, that the integral $(I,\pi_t)$ cannot be computed in closed form for this system. Therefore, we have used the ``expensive'' estimate
$$
(I,\pi_t) \approx (I,\pi_t^J), \quad \mbox{with $J=10^5$ particles},
$$
computed via the standard BF, as a proxy of the true value.

\subsection{Numerical results} \label{ssResults}

\begin{figure*}[htpb]
\centering
	\includegraphics[width=0.45\textwidth]{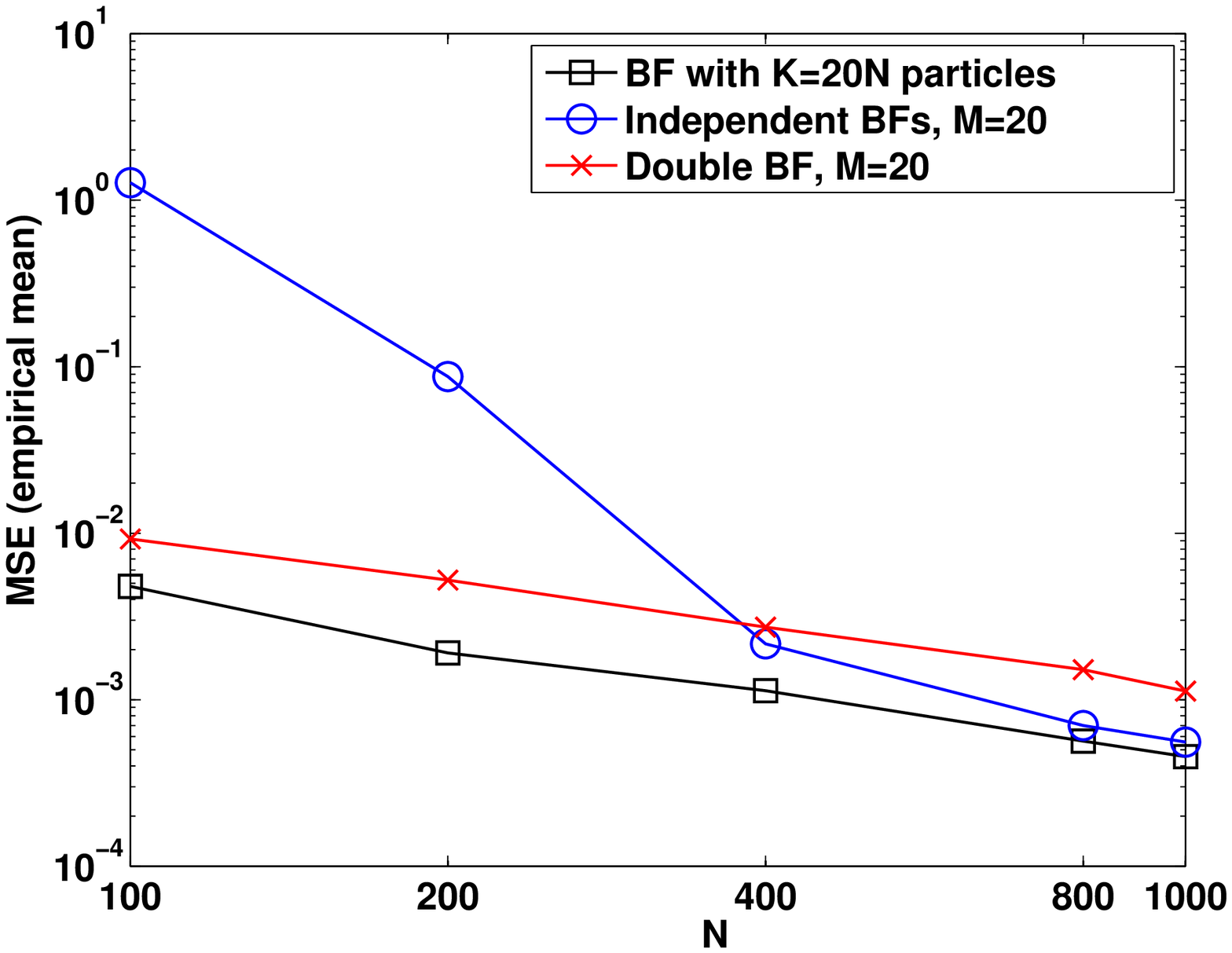} 
	\includegraphics[width=0.45\textwidth]{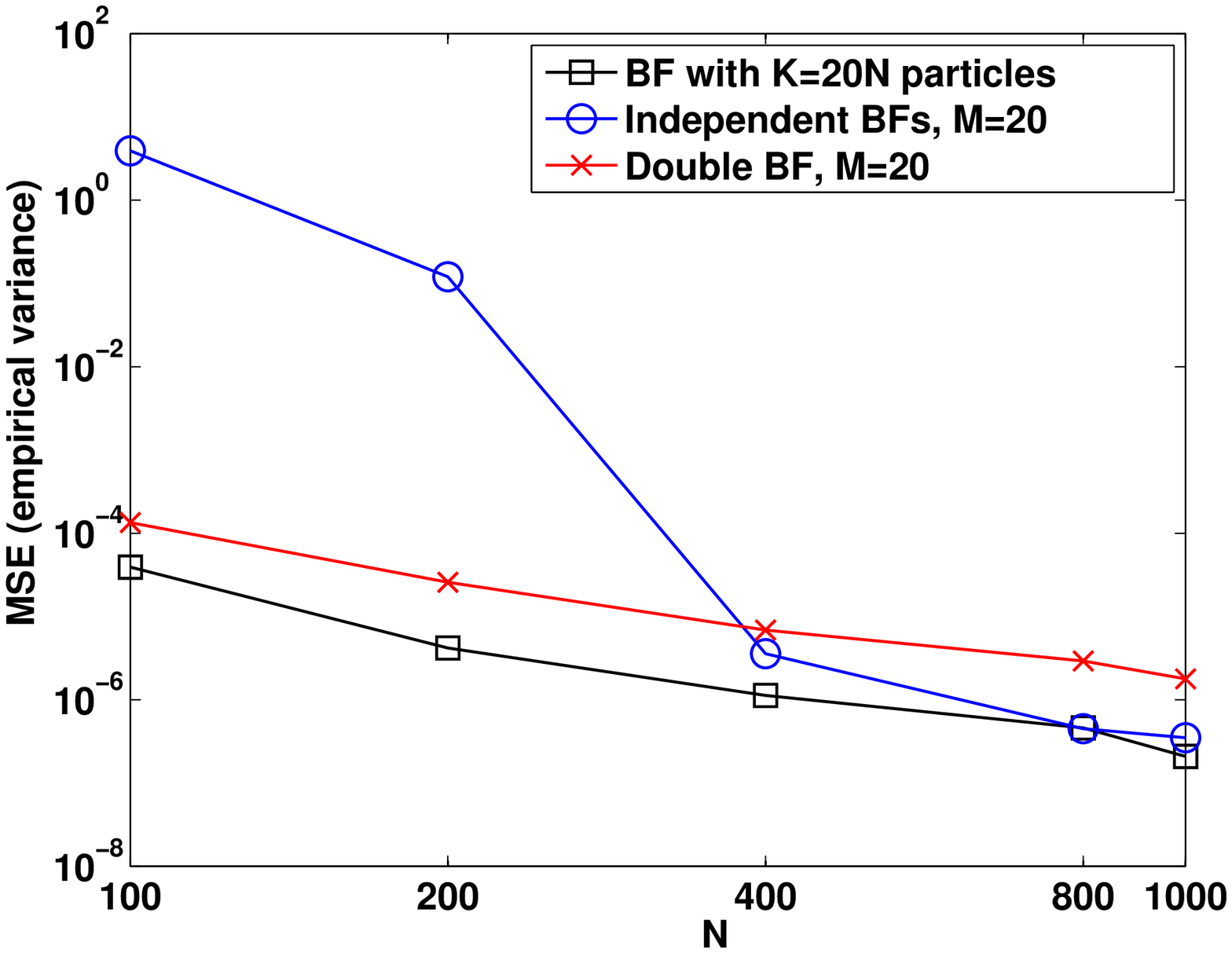} 
	\caption{Empirical mean (left) and variance (right) of the MSE for the centralised BF, the independent ensemble of BFs and the double BF method with $M=20$ constant and $N=100, 200, 400, 800, 1000$. All curves have been obtained from a set of 100 independent simulation trials.}
	\label{fMSE_fixedM}
\end{figure*}

Figure \ref{fMSE_fixedM} (left) displays the empirical MSE, averaged over 100 independent simulation runs, attained by the parallel schemes when the number of filters (respectively, particle islands for the double BF method) is fixed, $M=20$, and the number of particles per filter (particle island) ranges from $N=100$ to $N=1000$. The outcome of the centralised BF with $K=MN$ particles, hence ranging from $K=20\times 100$ to $K=20 \times 1000$, is also shown for comparison. We observe that proposed ensemble of independent BFs achieves a poor performance when the number of particles per filter, $N$, is relatively low ($N=100$), while for moderate values ($N \ge 400$) it nearly matches the MSE of the centralised BF. The double BF method is more accurate than the independent ensemble for $N=100, 200$, as it takes advantage of the interaction among the particle islands, but displays a slightly worse MSE than the centralised BF and the independent ensemble for $N \ge 400$. 
 
The empirical variance of the MSE for the same set of 100 simulation trials is displayed in Figure \ref{fMSE_fixedM} (right). The results show, again, that the double BF algorithm makes an efficient use of the island resampling step when $N$ is low, so that the $M$ islands remain balanced and the overall filter works properly, but falls short of the independent ensemble scheme for larger values of $N$.

%

Finally, we look into the relationship between the MSE and the running time for the three algorithms. With the number of filters (correspondingly, particle islands) $M=20$ fixed, we have run 100 independent simulation trials for each value $N=100, 200, 400, 800$ and $1000$, and computed the empirical MSE and the average running time for the two parallel schemes and each combination of $M$ and $N$. Correspondingly, we have also run the centralised BF with $K=MN$ particles, hence for $K=2 \times 10^3, 4 \times 10^3, 8 \times 10^3, 16 \times 10^3$ and $20 \times 10^3$. 

\begin{figure*}[htpb]
\centering
	\includegraphics[width=0.45\textwidth]{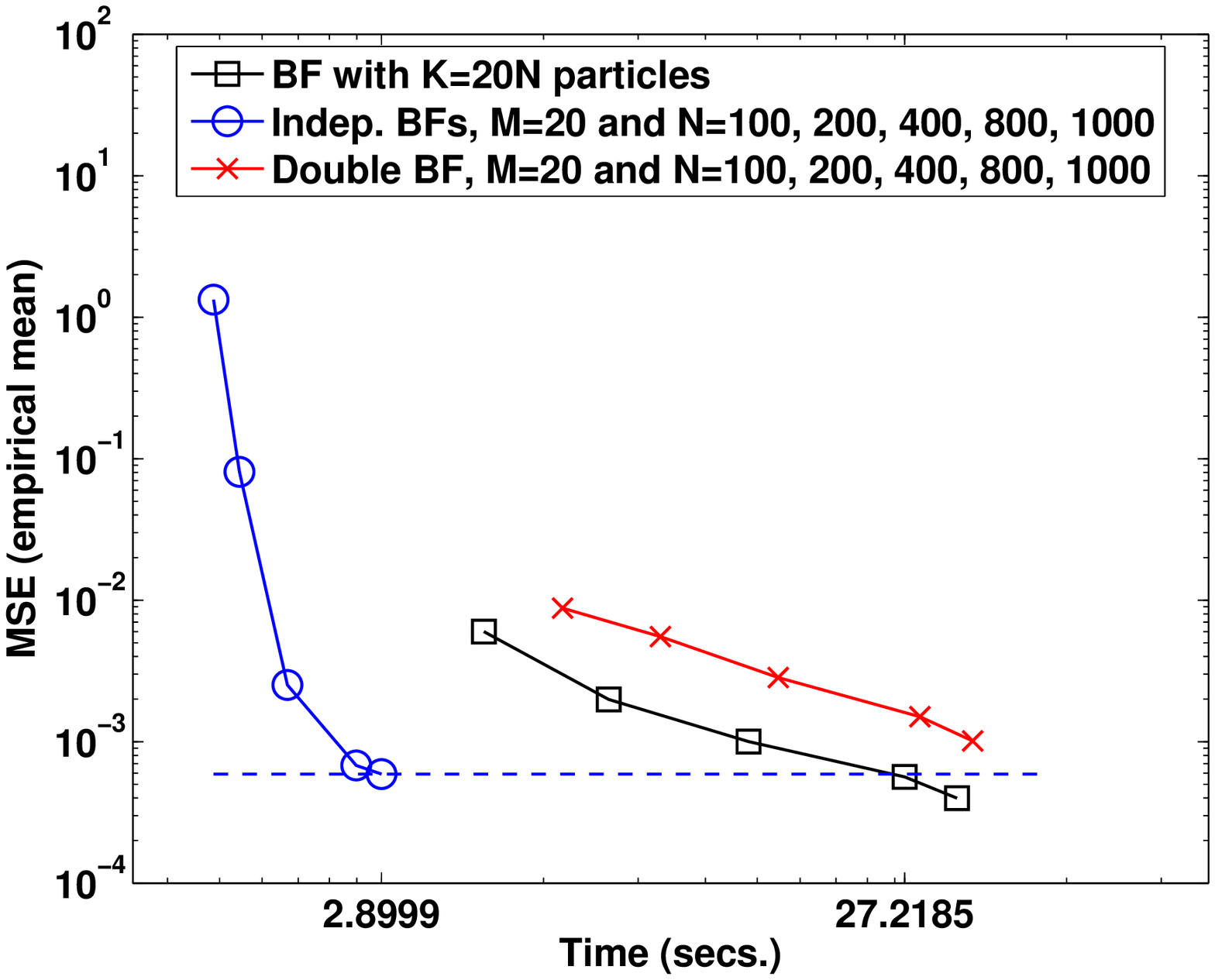} 
	\includegraphics[width=0.45\textwidth]{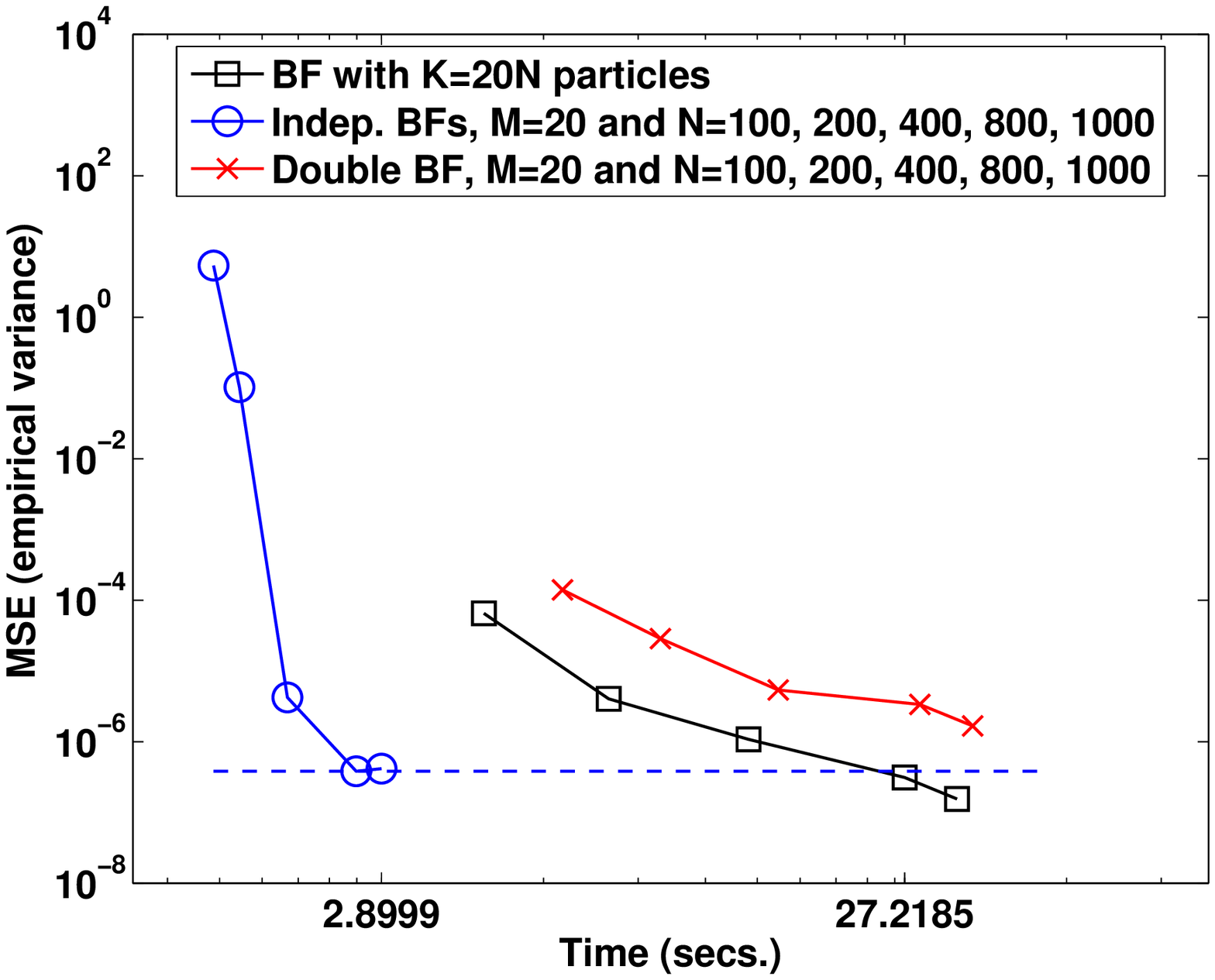} 
	\caption{Empirical mean (left) and variance (right) of the MSE versus the running time for the centralised BF, the independent ensemble of BFs and the double BF method. The two parallel schemes are run with $M=20$ constant and $N=100, 200, 400, 600, 800$ and $1000$. The centralised BF is run with $K=20N$ particles, where $N$ takes values in the same way as for the parallel algorithms. The dashed horizontal lines indicate where the mean (left) and variance (right) of the MSE match for the independent ensemble and the centralised BF. The running times for the two algorithms at that MSE level are shown as labels on the horizontal axis.}
	\label{fMSEvsTime}
\end{figure*}

Figure \ref{fMSEvsTime} (left) displays the resulting empirical MSE versus the running time for the three methods. If we qualify an algorithm as more efficient than another one when it is capable of attaining a lower MSE in the same amount of time, then this set of simulations shows that the independent ensemble scheme is more efficient than both the centralised BF and the double BF method. Indeed, a close look at Figure \ref{fMSEvsTime} (left) reveals that the ensemble of $M=20$ independent BFs with $N=1000$ particles per filter achieves an empirical MSE of $\approx 6 \times 10^{-4}$ with a running time of $\approx 2.9$ seconds, while the centralised BF attains the same performance with $K=20 \times 800$ particles and a running time of $\approx 27.2$ seconds (as shown by the dashed horizontal line in the plot). The double BF method falls short of this MSE value even with $M=20$ and $N=1000$. Figure \ref{fMSEvsTime} (right) shows the empirical variance of the MSE, versus the running time, for the same set of computer simulations\footnote{These results show that a straightforward implementation of the double BF in Matlab, to be run on a multicore server, is not particularly efficient, because of the overhead due to the subset-level resampling. We do not imply that other implementations of this algorithm (e.g., using GPUs, FPGAs or simply a different programming language) should be equally inefficient.}.

\section{Example: A stochastic dynamical network} \label{sFHN}

%
\subsection{State space model} \label{ssModel_fhn}

As a second example we have studied a dynamical complex model consisting of a network of {\em modified} stochastic FitzHugh-Nagumo (FH-N) nodes. This model is designed to mimic some of the dynamical patterns that can be observed in the electrical activity of the human heart during atrial fibrillation episodes (see \cite{Keener08a} for a survey of the topic). Let us consider a 2-dimensional rectangular grid consisting of $J \times J$ nodes, where each node is a stochastic dynamical subsystem that can be described by the classical FitzHugh-Nagumo equations \cite{Keener08a} plus 
\begin{itemize}
\item a stochastic (additive noise) term,
\item a random stimulus, and 
\item a coupling term that determines the interaction with the neighbour nodes. 
\end{itemize}
To be specific, the continuous-time dynamics of the node in the $(i,j)$ position of the grid, with $1 \le i \le J$ and $1 \le j \le J$, follows the stochastic differential equation
\begin{eqnarray}
dU_{i,j} &=& \left(
	p_3(U_{i,j}) - V_{i,j} + \frac{1}{D} \sum_{(l,r) \in \mN_{i,j}} U_{l,r} \right)dt \nonumber \\
&&+ \left(m_{i,j} F + \Psi_{i,j} \right)dt + \tilde\sigma dB_{i,j}, 
\label{eqU}
\end{eqnarray}
where: 
\begin{itemize}
\item the continuous-time stochastic process $\{ U_{i,j}(s) \}_{s \ge 0}$ usually represents an action potential (or any other voltage-related signal) in biological models \cite{Keener08a}; 
\item $p_3(u) = \sum_{r=0}^3 \alpha_r u^r$ is a polynomial of order 3 with known fixed coefficients $\alpha_r$, $r=0, 1, 2, 3$; 
\item $\{ V_{i,j}(s) \}_{s \ge 0}$ is the so-called recovery process, that evolves according to the differential equation
\begin{equation}
dV_{i,j} = \beta_0 U_{i,j} + \beta_1 V_{i,j} + \beta_2 ds
\label{eqV}
\end{equation}
with known parameters $\beta_r$, $r=0, 1, 2$;
\item the set $\mN_{i,j} \subset \{ 1, ..., J \} \times \{ 1, ..., J \}$ contains the neighbours, in the grid, of the $(i,j)$-th node;
\item the coupling coefficient $D>0$ is known and fixed;
\item $F(s) : (0,\infty) \rw \Real^+$ is a known, typically periodic, forcing signal;
\item $m_{i,j} \in \{ 0, 1 \}$ is a (known and fixed) binary indicator that determines which nodes are excited by the forcing signal $F(s)$; 
\item $\Psi_{i,j}(s)$ is a random stimulus that can be applied to nodes which are returning to a state of quiescence\footnote{The specific random procedure producing this stimulus is described in Section \ref{ssSetup_fhn}. The goal is to allow the model to generate re-entries of excitation waves which, in turn, lead to stable and self-sustained dynamical patterns of different classes.}, and
\item $\{ B_{i,j}(s) \}_{s \ge 0}$ are standard and independent Wiener processes and the scale parameter $\tilde \sigma$ is assumed known.
\end{itemize}

For the simulations in this section we have obtained simple discrete-time versions of \eqref{eqU} and \eqref{eqV} using Euler's method with an integration time step $T_d$, which yields  
\begin{eqnarray}
U_{i,j,t} &=& U_{i,j,t-1} + T_d\left(
	p_3(U_{i,j,t-1}) - V_{i,j,t-1} \right) \nonumber \\	
&&+ T_d\left( \frac{1}{D} \sum_{(l,r) \in \mN_{i,j}} U_{l,r,t-1} + m_{i,j} F_t + \Psi_{i,j,t}
\right) \nonumber \\
&&+ \tilde \sigma \sqrt{T_d} B_{i,j,t}, \label{eqUd} \\
V_{i,j,t} &=& V_{i,j,t-1} + T_d \left(
	\beta_0 U_{i,j,t-1} + \beta_1 V_{i,j,t-1} + \beta_2 
\right), \label{eqVd}
\end{eqnarray}
where $t = 1, 2, ...$ denotes discrete time, $\{ U_{i,j,t} \}_{t=0, 1, ...}$ is the signal sequence at the $(i,j)$-th node, $\{ V_{i,j,t} \}_{t=0, 1, ...}$ is the recovery sequence at the same node, $\{\Psi_{i,j,t}\}_{t=1,2,...}$ is a sequence of random stimuli and $\{ B_{i,j,t} : 1 \le i \le J, 1 \le j \le J, t \ge 0 \}$ is a set of i.i.d. Gaussian random variables with zero mean and unit variance. 

The interest in this model is to study dynamical patterns in the network described by Eqs. \eqref{eqUd}-\eqref{eqVd}, which are caused by the sequence of random stimuli ${\bf \Psi}_t = \{ \Psi_{i,j,t} : 1\le i \le j, 1 \le j \le J \}_{t \ge 0}$. It is known that when a sufficiently strong stimulus (in the form of a positive shift of the voltage signal $U_{i,j,t}$) is applied to the network in a region that had been recently excited by the periodic forcing, the propagation of this stimulus can lead to a re-entrant wavefront that becomes stable and can prevail over the excitations caused by the forcing signal $F(s)$ (see, e.g., \cite{Keener08b} for models related to cardiac tissue). 

We consider the following model for the sequence ${\bf \Psi}_t$. For each $t \ge 1$ and a pair of given thresholds ${\sf u}_- < {\sf u}^+$, we identify the stimulation region
\begin{eqnarray}
\kS_t :=  \left\{
	(i,j) \in \{ 1, ..., J \}^2 : {\sf u}_- < U_{r,k,t-1} < {\sf u}^+  \right. \nonumber \\
\left. \mbox{ for some } (r,k)\in\{(i,j)\} \cup \mN_{i,j} \right\},
\label{eqStimulusRegion}
\end{eqnarray}
i.e., $\kS_t$ consists of the nodes $(i,j)$ such that the state of the node at time $t-1$ lies between the thresholds, $U_{i,j,t-1} \in ({\sf u}_-,{\sf u}^+)$, {\em or} the state of some neighbour at time $t-1$, $(r,k) \in \mN_{i,j}$, lies between the thresholds, $U_{r,k,t-1} \in ({\sf u}_-,{\sf u}^+)$. At each time step, a new stimulus is applied to a group of neighbour nodes in the activation region $\kS_t$ with (a small) probability $\epsilon \ge 0$. To be specific, let  
\begin{equation}
{\sf B}_t^\epsilon \sim \mbox{Bernoulli}(\epsilon), \quad t = 1, 2, ...,
\label{eqBernu}
\end{equation}
be a sequence of i.i.d. Bernoulli random variables with parameter $0 \le \epsilon <1$ and, for each $(i,j) \in \{1, ..., J \}^2$, let
\begin{equation}
(i_t^*, j_t^*) \sim {\sf Uniform}(\kS_t)
\label{eqUnifStimulus}
\end{equation}
be a single pair of indices drawn from the uniform distribution with support on the stimulus region $\kS_t$. Then,  
we can generate a random indicator $Q_{i,j,t}^\epsilon$ of the form
\begin{equation}
Q_{i,j,t}^\epsilon = \left\{
	\begin{array}{cl}
	0, &\mbox{if $(i,j) \notin \{(i_t^* ,j_t^*)\} \cup \mN_{i_t^*,j_t^*}$}\\
	{\sf B}_t^\epsilon &\mbox{if $(i,j) \in \{(i_t^* ,j_t^*)\} \cup \mN_{i_t^*,j_t^*}$}\\
	\end{array}
\right.
\label{eqQij}
\end{equation}
which selects a set of neighbouring nodes where a new stimulus is to be applied at time $t$. Then, the sequence of stimuli in the $(i,j)$ node can be written as
\begin{equation}
\Psi_{i,j,t} = \tilde F \max\left\{ 
	1, \sum_{l=0}^{\ell_0-1} Q_{i,j,t-l}^\epsilon
\right\},
\label{eqPsi_ijt}
\end{equation}
where $\tilde F$ is the amplitude of each single stimulus, which is sustained during $\ell_0$ consecutive discrete-time steps.

Let us denote
\begin{eqnarray*}
	X_{i,j,t} = ( U_{i,j,t}, V_{i,j,t}, Q_{i,j,t-\ell_0+1:t} ) \in \Real^{2+\ell_0} \quad \mbox{and} \\ 
	X_t = ( X_{1,1,t}, \ldots, X_{1,J,t}, \ldots, X_{J,J,t} ) \in \Real^{(2+\ell_0)J^2}.
\end{eqnarray*}

The $(2+\ell_0)J^2$-dimensional sequence $\{ X_t \}_{t \ge 0}$ is a Markov process in discrete time. We do not attempt to write down the associated transition kernel $\tau_t(dx|x_{t-1})$ explicitly, yet it is straightforward to generate a sample $X_t$ conditional on $X_{t-1}=x_{t-1}$ using Eqs. \eqref{eqStimulusRegion}--\eqref{eqPsi_ijt}, \eqref{eqUd} and \eqref{eqVd}, in this particular order, over the set of indices $\{ (i,j) : 1 \le i \le J, 1 \le j \le J \}$.

To complete the state-space model description, we assume the ability to observe the signal (voltage) variables, $U_{i,j,t}$, in a subset of the nodes of the grid. To be specific, at time $t$ we collect the measurements
\begin{equation}
Y_{i,j,t} = U_{i,j,t} + \bar \sigma \bar B_{i,j,t}, \quad (i,j) \in \mS_y, 
\label{eqObservFHN}
\end{equation}
where $\mS_y \subset \{ 1, ..., J \} \times \{ 1, ..., J \}$ is the set of observed nodes, $\{ \bar B_{i,j,t} : (i,j) \in \mS_y \}$ is a set of i.i.d. standard Gaussian random variables (centred and with unit variance) and $\bar \sigma^2$ is a known scale parameter. The full observation at time $t$ is denoted
$$
Y_t = \{ Y_{i,j,t} : (i,j) \in \mS_y \} \in \Real^{|\mS_y|}.
$$  
The likelihood function is Gaussian, namely
\begin{equation}
g_t^{y_t}(x_t) \propto \exp\left\{
	-\frac{1}{2\bar \sigma^2} \sum_{(i,j) \in \mS_y} \left(
		y_{i,j,t} - u_{i,j,t}
	\right)^2
\right\}.
\nonumber 
\end{equation} 

Equations \eqref{eqStimulusRegion}--\eqref{eqPsi_ijt}, \eqref{eqUd}, \eqref{eqVd} and \eqref{eqObservFHN} describe a Markov state--space model in discrete time, with conditionally independent observations. We aim at tracking the sequence of probability measures 
$$
\pi_t(A) = \Prob\{ X_t \in A | Y_{1:t} = y_{1:t} \}, \quad t=1, 2, ..., T,
$$ 
where $A \in \mB(\mX)$ and $y_{1:T}$ is a given sequence of observations.

%
\subsection{Simulation setup} \label{ssSetup_fhn}

We have run simulations for a network of $J^2=32^2=1,024$ modified stochastic FH-N nodes, interconnected in a regular square grid. Therefore, for an ``inner'' node $(i,j) \in \{ 1, ..., J \}^2$ with $1 < i, j < J$ the set of neighbours is 
$$
\mN_{i,j} = \left\{ (r,l) \in \{ 1, ..., J \}^2 : r = i \pm 1, l = j \pm 1 \right\},
$$
whereas for the nodes in the ``corners'' of the grid the sets of neighbours are
\begin{eqnarray}
\mN_{1,1} &=& \{ (1,2), (2,1) \}, \nonumber \\
\mN_{1,J} &=& \{ (1,J-1), (2,J) \}, \nonumber \\
\mN_{J,1} &=& \{ (J,2), (J-1,1) \}, \nonumber \\
\mN_{J,J} &=& \{ (J-1,J), (J,J-1) \}, \nonumber
\end{eqnarray}
and for the nodes on the ``sides'' of the grid
\begin{eqnarray}
\mN_{1,j} &=& \{ (1,j\pm 1), (2,j) \}, \nonumber\\
\mN_{j,1} &=& \{ (j \pm 1, 1), (j,2) \}, \nonumber\\
\mN_{j,J} &=& \{ (j \pm 1, J), (j,J-1) \}, \nonumber\\
\mN_{J,j} &=& \{ (J, j \pm 1), (J-1,j) \},\nonumber
\end{eqnarray}
where $1 < j < J$ in all cases. The time discretisation period is $T_d=5\times 10^{-3}$ continuous-time units and the coupling constant, that sets the ``strength'' of the links between neighbours, is $\frac{1}{D}=4.5 \times 10^{-3}$. 

The dynamics of the FH-N system is highly dependent on the choice of the polynomial $p_3(u)$ in Eq. \eqref{eqUd}, which for this set of simulations is selected as
$$
p_3(u) = u\left(
	u + \sqrt{\frac{18}{5}}
\right)\left(
	u - \sqrt{\frac{18}{5}}
\right),
$$
and the forcing signal $F(s)$, which hereafter consists of a periodic sequence of pulses of the form
$$
F(s) = \sum_{k=0}^\infty \sqcap(s - kS_\sqcap),
$$
where $\sqcap(s)$ is the square waveform
$$
\sqcap(s) = \left\{
	\begin{array}{cl}
	\tilde F &\mbox{if } 0 \le s \le S_\sqcap\\
	0 &\mbox{otherwise}\\
	\end{array}
\right.,
$$
the period of $F(s)$ is $S_\sqcap = 20$ time units and the amplitude of the pulses is $\tilde F = 200$. The discrete-time forcing signal is $F_t = F(s=tT_d)$. 

To construct the stimulus region $\kS_t$ given by Eq. \eqref{eqStimulusRegion} we use the thresholds
$$
{\sf u}_- = -1.8 \quad \mbox{and} \quad {\sf u}^+ = -1.6,
$$
which correspond to the back tail of a wave propagating over the network. The parameter of the Bernoulli distribution in \eqref{eqBernu} is $\epsilon = 10^{-3}$, the amplitude of the stimulus in \eqref{eqPsi_ijt} is $\tilde F = 200$ and it is sustained for $\ell_0=25$ discrete time steps. The variance of the noise term in Eq. \eqref{eqUd} is $\tilde \sigma^2 = \frac{1}{2}$ and the specification of the dynamics is complete with the parameters $(\beta_0,\beta_1,\beta_2) = (2.1,-0.6,0.6)$ in Eq. \eqref{eqVd}.

\begin{figure}[htb]
\centering
\includegraphics[width=0.8\linewidth]{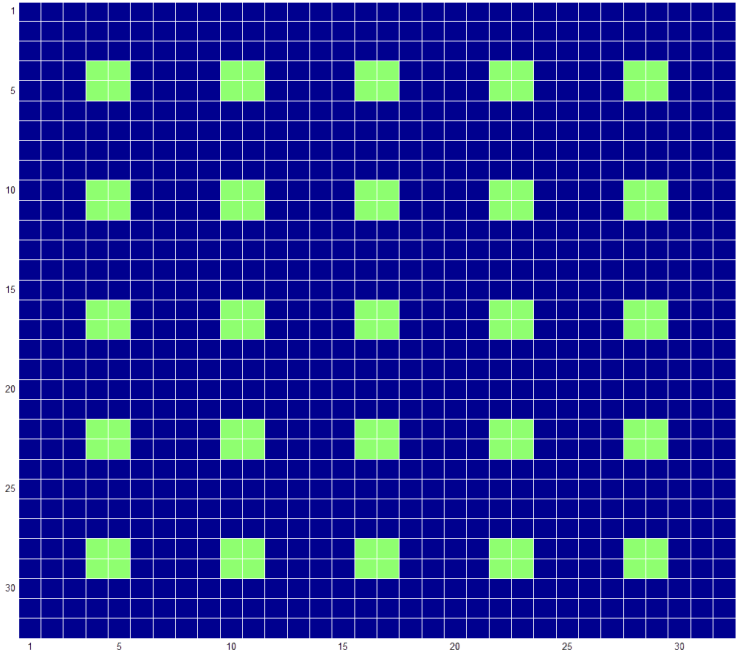}
\caption{Location of the observations within the $32 \times 32$ grid of stochastic FH-N nodes with random stimuli. The blue squares represent the nodes in the network, each one following the dynamics specified by Eqs. \eqref{eqStimulusRegion}--\eqref{eqPsi_ijt}, \eqref{eqUd} and \eqref{eqVd}. Observed nodes are displayed in green. Each observation zone includes four nodes, forming a $2 \times 2$ square. The set $\mS_y$ in \eqref{eqObservFHN} contains exactly the green-coloured nodes in this figure.} 
\label{fObservGrid}
\end{figure}

The observations are collected at a grid of $5 \times 5$ equally-spaced zones, each zone consisting of four nodes forming a $2 \times 2$ square, as shown in Figure \ref{fObservGrid}. Therefore, we collect observations from $5 \times 5 \times 2 \times 2 = 100$ nodes out of $J^2= 1,024$ in the network. For each observed node, say in the position $(r,l)$, we obtain the measurement specified by Eq. \eqref{eqObservFHN}, where the noise variance is $\bar \sigma^2 = \frac{1}{2}$.

%
\subsection{Numerical results} \label{ssResults_fhn}

We have run computer simulations for this model using Matlab version R2012b (32 bits), with the parallel computing toolbox enabled, on an a multicore Intel(R) Xeon(R) CPU E5-2680 v2 @ 2.80GHz server.  All the results reported here are based on a set of 20 independent forward simulations of the stochastic FH-N network model described in Sections \ref{ssModel_fhn} and \ref{ssSetup_fhn}. The state trajectories for these simulations have been recorded and synthetic observations have been generated from them according to \eqref{eqObservFHN}.

\begin{figure*}[htb]
	\includegraphics[width=\textwidth]{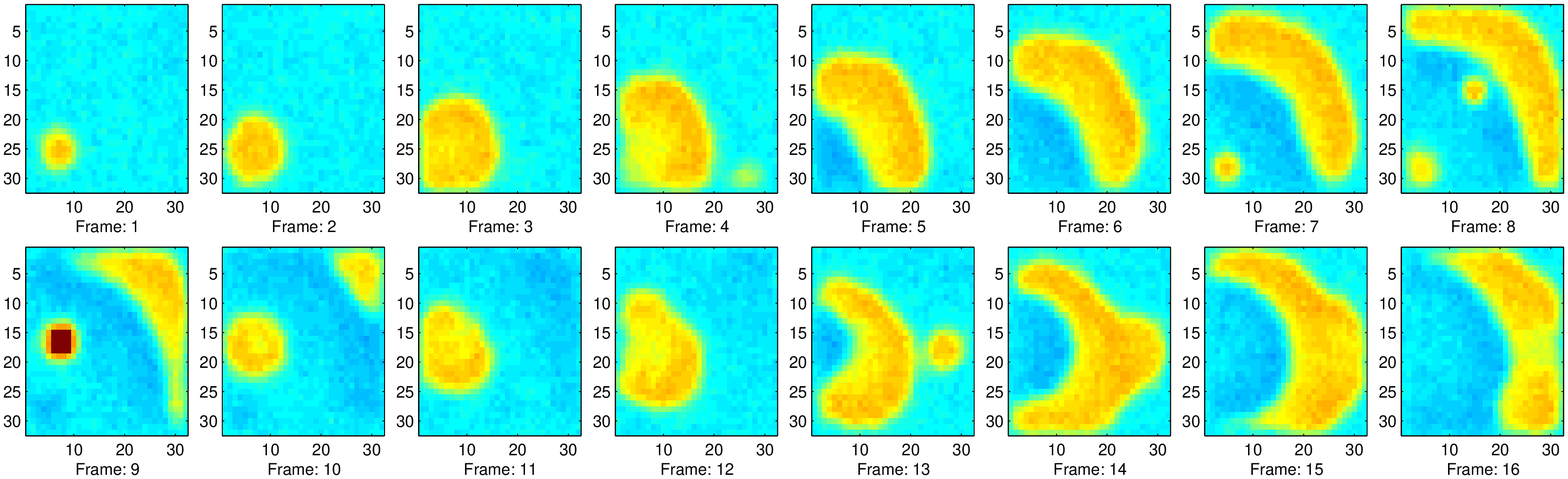}
	\includegraphics[width=\textwidth]{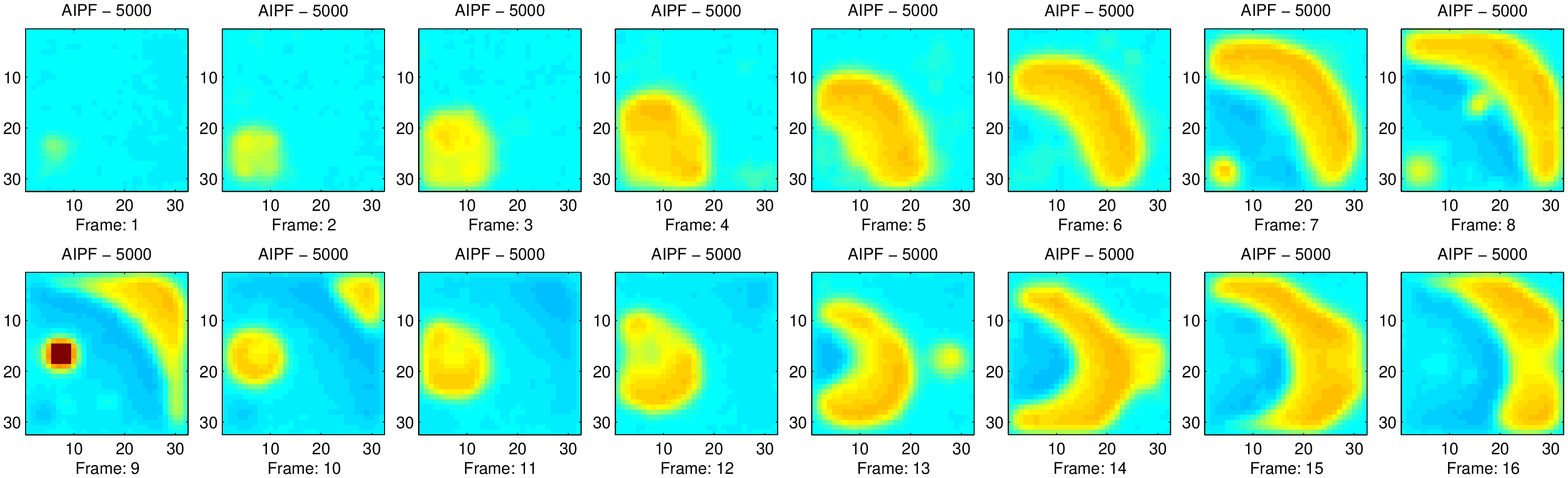}
	\caption{\textbf{Top:} Example of a wave propagating over the 32x32 grid, where hotter colours describe higher action potential $U_{i,j,t}$ values while colder ones represent nodes at their rest state. A stimulus is applied in the first frame and then propagates through the network creating a wave. Random stimuli behind this wave appear in frames 7, 8, 9 and 13. The stimulus that appears in frame 9 is strong enough to initiate another wavefront that propagates from that location, hence disorganising the periodic behaviour induced by the forcing signal $F_t$. \textbf{Bottom:} Posterior-mean estimation of the process $U_t$ using an ensemble of $M=10$ independent auxiliary particle filters (labeled AIPF) with $N=5,000$ particles each. The filter detects the initial stimulus and tracks the resulting wave. It also ``catches'' the random stimuli in frames 7, 8, 9 and 13, and tracks the wavefront initiated by the stimulus in frame 9.}
	\label{fPropagationExample}
\end{figure*}

Figure \ref{fPropagationExample} (top) displays the actual variables $U_{i,j,t}$ for $i, j = 1, ..., 32$ (i.e., the complete network) at different time steps for one of the simulations. This is a typical realisation of the process $U_t$, with a wave propagating over the network and occasional stimuli appearing behind the wave. The random stimulus in frame 9 is strong enough to start a wavefront that propagates over the network and moves the system away from the periodic behaviour that would be induced by the forcing signal $F_t$ alone. 

In order to track $U_t$ (or, indeed, the whole state $X_t$) we have applied an ensemble of independent auxiliary particle filters \cite{Pitt01}, which are labeled AIPF in Figure \ref{fPropagationExample} (bottom). In particular, we have run $M=10$ independent filters with $N=5,000$ particles each and the figure shows how the algorithm successfully detects the first stimulus, follows the resulting wavefront and, later, detects the various random stimuli (in frames 7, 8, 9 and 13) and also tracks the wave started by the stimulus in frame 9.

\begin{figure*}[htb]
\centerline{
\includegraphics[width=0.45\textwidth]{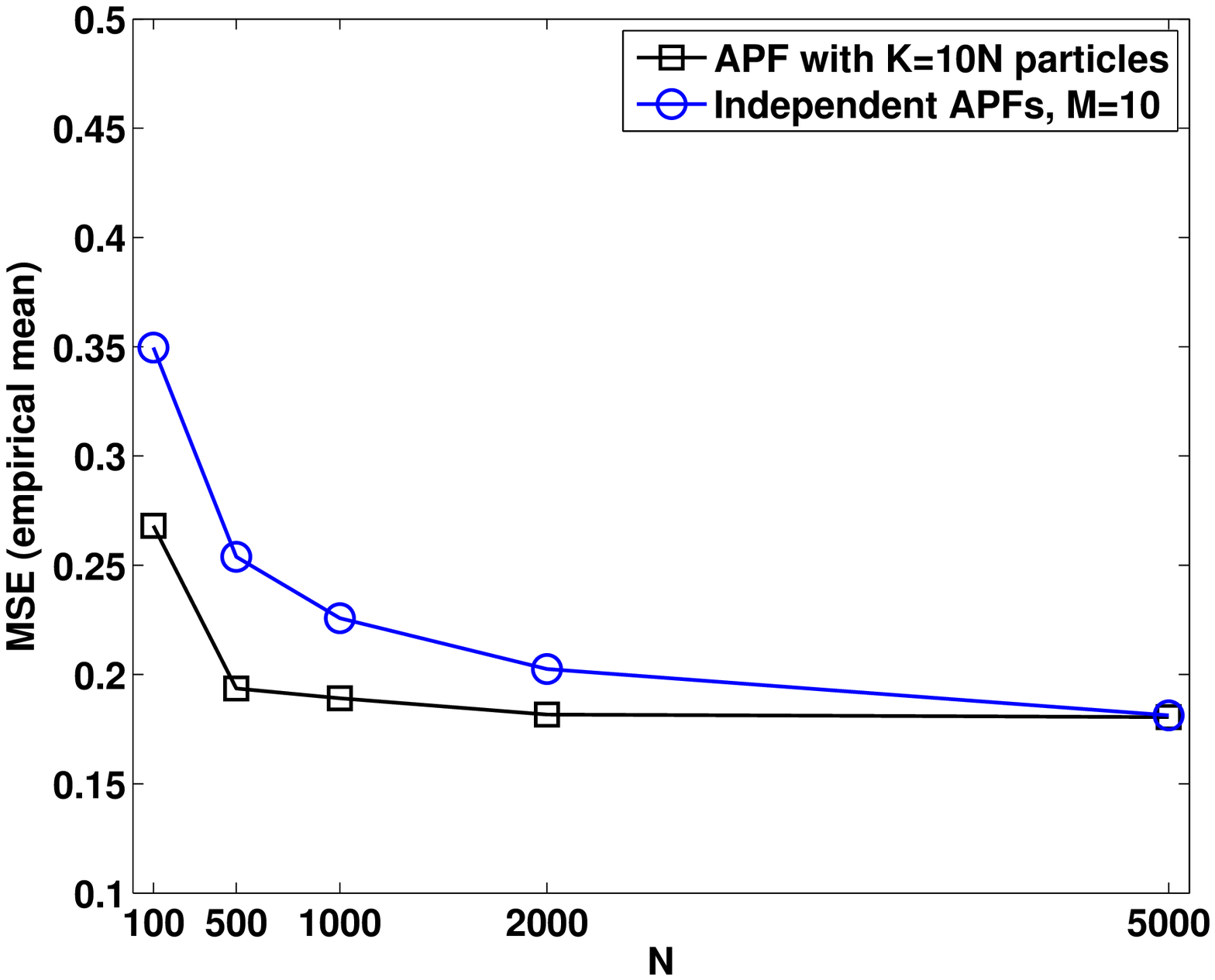} 
\includegraphics[width=0.45\textwidth]{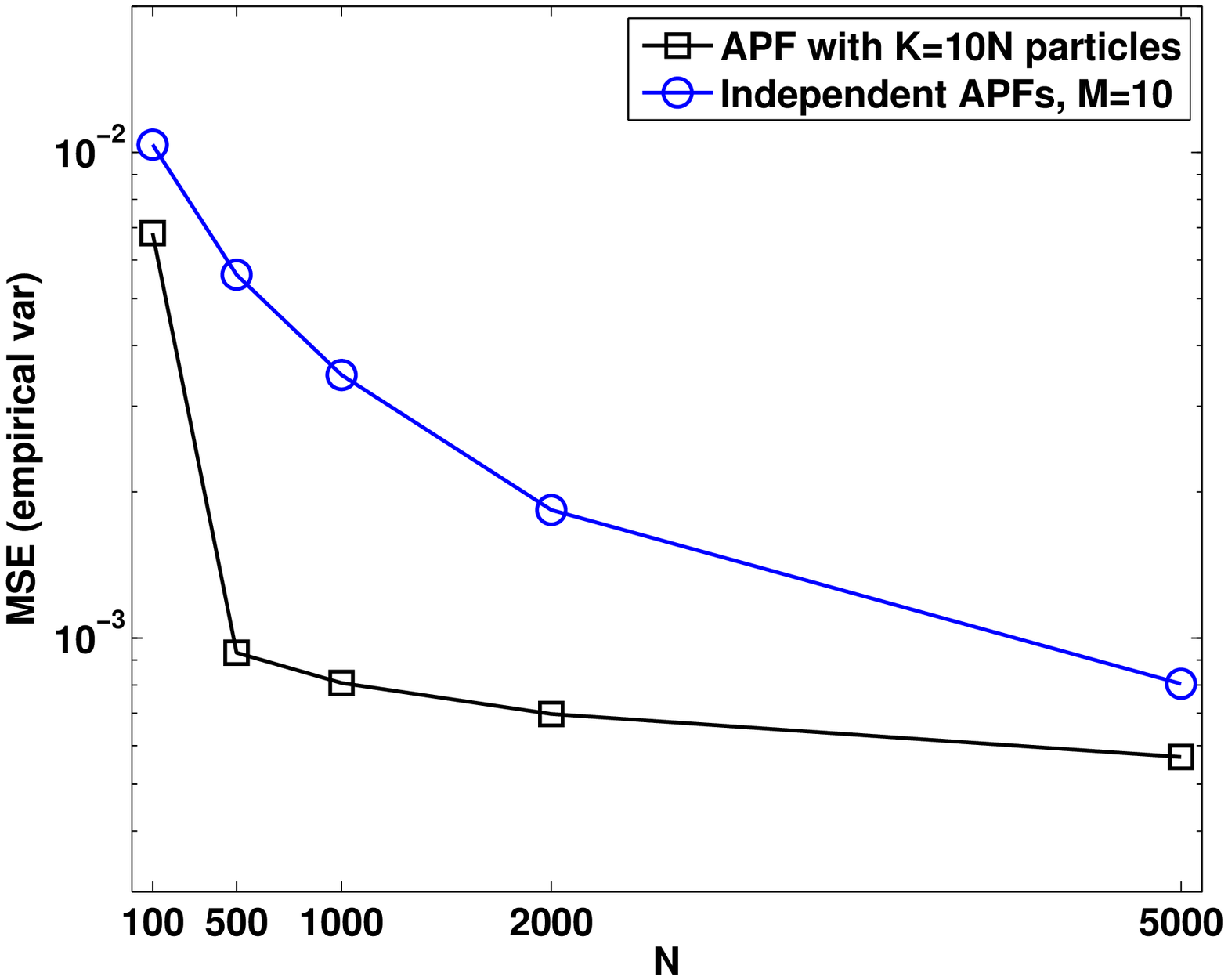} 
}
\caption{\textbf{Left:} Empirical MSE (per node) versus the number of particles per independent filter, $N$, averaged over 20 independent simulation runs. The results correspond to an ensemble of $M=10$ independent auxiliary particle filters. \textbf{Right:} Empirical variance of the MSE (per node) versus the number of particles per independent filter, $N$. The performance of the centralised auxiliary particle filter with $K=MN$ particles is also plotted, for reference, in the two plots.} 
\label{fMSEvsN_fhn}
\end{figure*}

Figure \ref{fMSEvsN_fhn} (left) shows the empirical MSE per node of the state estimates generated by different ensembles of auxiliary particle filters, namely $M=10$ independent filters with the number of particles per filter varying from $N=100$ to $N=5,000$. The same MSE for the centralised auxiliary particle filters with $K=MN$ particles (i.e., with $K=1,000, \ldots, 50,000$) is also plotted for reference. The same as for the Lorenz 63 example, we observe that the relative performance of the ensemble of independent filters is poor for smaller values of $N$, yet for $N \ge 2,000$ it nearly matches the MSE of the centralised particle filter. The results have been averaged over the 20 independent data sets described at the beginning of this Section. The empirical variance of the MSE per node for the same set of simulations is depicted in Figure \ref{fMSEvsN_fhn} (right).

\begin{figure*}[htb]
\centerline{
\includegraphics[width=0.45\textwidth]{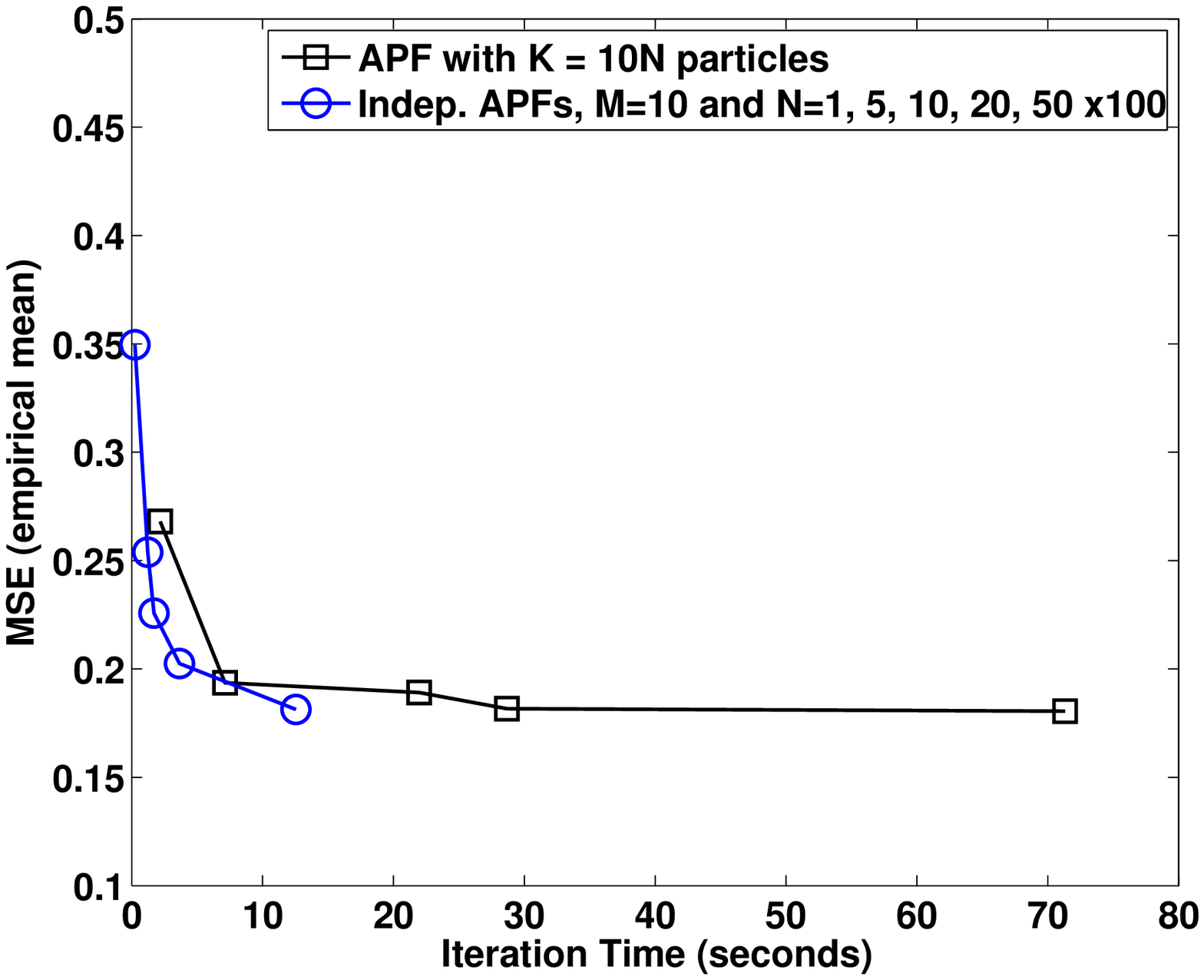} 
\includegraphics[width=0.45\textwidth]{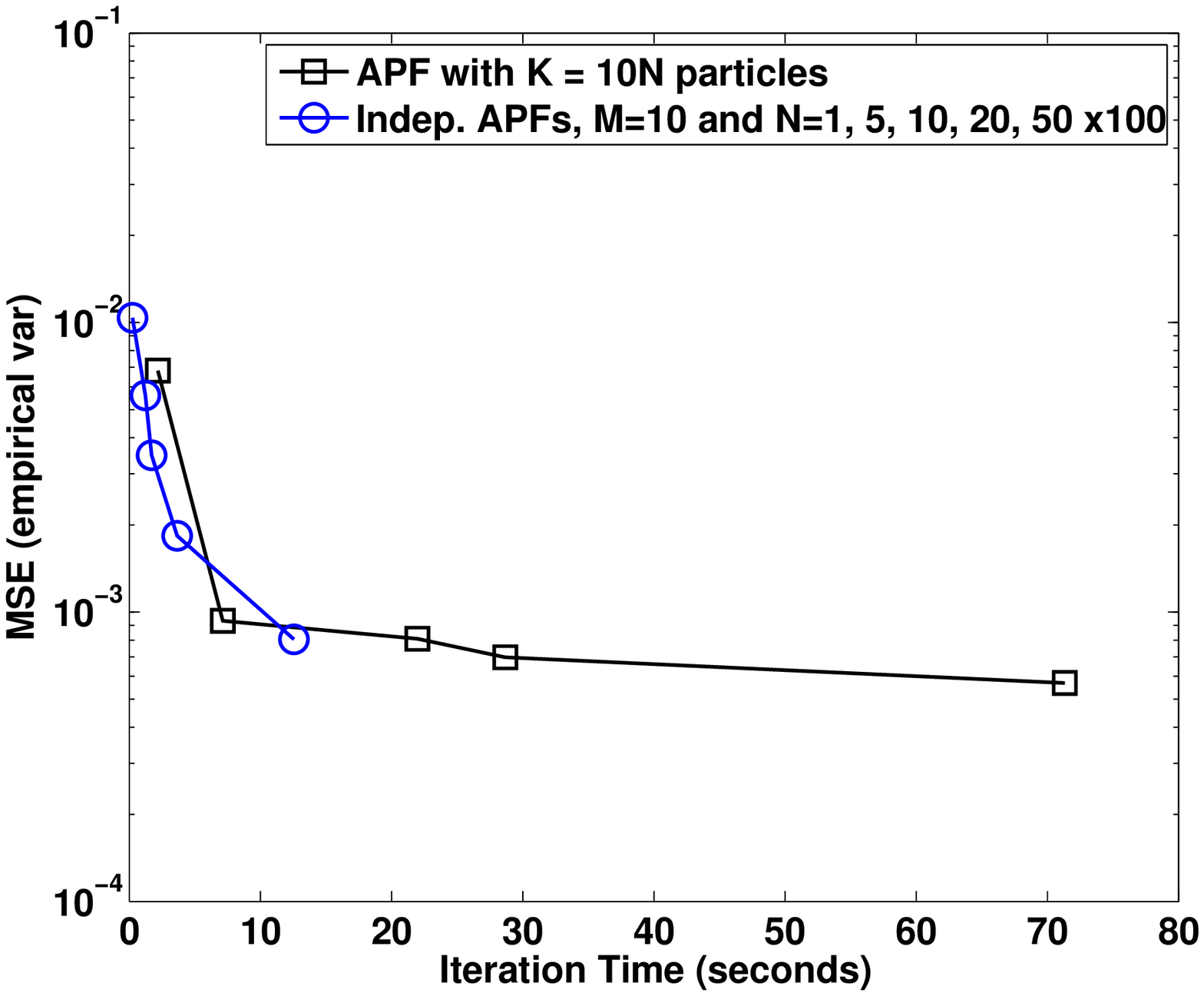} 
}
\caption{\textbf{Left:} Empirical MSE (per node) versus running time (per discrete time step), averaged over 20 independent simulation runs. Each point in the curves corresponds to a different value of $N$, namely $N = 100, 500, 1000, 2000$ and $5000$. The centralised auxiliary particle filter is run with $K=10\times N$ particles, for the same values of $N$. \textbf{Right:} Empirical variance of the MSE (per node) versus the running time (per discrete time step) for the same set of simulations.} 
\label{fMSEvstime_fhn}
\end{figure*}

Finally, Figure \ref{fMSEvstime_fhn} (left) displays the empirical MSE (per node) versus the running time (per discrete time step) for an ensemble of $M=10$ independent auxiliary particle filters and for the centralised auxiliary particle filter with $K=MN$ particles. For the ensemble of independent filters, each point in the curve corresponds to a different value of $N$ (particles per filter), and for the centralised filter, each point corresponds to a different value of $K=10N$, the total number of particles. In particular, we have obtained results for $N = 100, 500, 1000, 2000$ and $5000$.  A close look at the figure shows that the ensemble of $M = 10$ independent filters with $N = 5,000$ particles each achieves an empirical MSE of  $\approx 0.1813$ with a running time of  $\approx 12.56$ seconds, while the centralised particle filter attains the same MSE with $K=10 \times 2,000$ particles and a running time of  $\approx 28.67$ seconds (no improvement is observed for $K=10 \times 5,000$ particles, while the running time scales up to $\approx 73$ seconds). Therefore, the ensemble of independent filters turns out more efficient than the centralised algorithm for this example as well.

\section{Discussion} \label{sConclusions}

%
%
%

We have addressed the problem of parallelising the standard particle filtering algorithm\footnote{While we have restricted the research to the standard (bootstrap) filter for simplicity of presentation, the analysis, and the whole argument about parallelisation by means of independent filters, extends in an almost straightforward manner to more sophisticated algorithms that may use tailored importance functions and/or auxiliary variables for the generation of particles.} by splitting the total number of particles $K$ into $M$ subsets, running one independent particle filter per subset, and then building an average filter measure using the ensemble of i.i.d. random approximations produced by the filters. This approach avoids all dependences among the filters, an idea which goes against recent approaches to the problem \cite{Bolic05,Whiteley13,Verge13,Miguez15,Heine15} and, to some extent, against the intuition that a certain interaction is needed to make the $M$ filters, with $N$ particle each, work together with the same performance as a centralised filter with $K=MN$ particles. 

The~rationale~to~advocate~the averaging of independent particle filters instead of the (algorithmically more sophisticated) schemes based on controlled interactions, such as \cite{Bolic05,Whiteley13,Verge13}, relies on both theoretical and practical arguments. Theoretically, the simple analysis in this paper shows that the parallelisation based on independent filters has the same asymptotic performance as a centralised particle filter. We have obtained this result by looking at the mean square approximation error for integrals of bounded functions w.r.t. the filter measure, decomposed into variance and bias terms. The bounds we have obtained depend explicitly on $M$ (the number of filters) and $N$ (the number of particles per filter), and show that there is no asymptotic performance loss for schemes with $N \ge M$. This result is actually aligned with recent contributions in the field of machine learning regarding the statistical properties of averages of independent estimators (including classifiers and regressors) \cite{Rosenblatt14}. We have also utilized the asymptotic convergence rates to propose a time--error index that enables a quantitative comparison of centralised and parallel particle filtering scheme in terms of asymptotic accuracy and running time. These analytical results can be extended to account for stronger forms of convergence (under additional assumptions on the model, see \cite{Heine15}) and adapted to continuous-time state-space systems (see \cite{Han13thesis} for the analysis of the bias and the MSE in this context). 

From a practical perspective, we have shown that the averaging of independent filters should be preferred when $N$, the number of particles per independent filter, is sufficiently large. Indeed, our computer simulations suggest that if we seek a scheme with a large number of parallel filters ($M$) and a relatively small number of particles per filter ($N$) then parallelisation schemes that exploit a certain degree of interaction between filters should be preferred --at the expense of a computational overhead to implement such interaction. On the other hand, if $N$ is large enough to make the parallel filters work (even roughly), then our simulations show that interaction is not needed anymore and independence can be fully exploited both in terms of accuracy 
and running time. 

The interest in designing particle filtering schemes that can have fast implementations using massively parallel hardware has followed the surge of several problems in science (geophysics, biochemistry or systems medicine) and engineering (sensor networks, multi target tracking) where the fundamental task is the tracking of a complex, high-dimensional dynamical system. In this paper we have investigated an example where the system of interest is a network of 1,024 stochastic FitzHugh-Nagumo nodes with random stimuli, interconnected through a 2-dimensional regular grid. This kind of complex stochastic network is a good representative of models commonly used for excitable media in ecology or medicine \cite{Keener08a} and illustrates the kind of models for which parallel particle filters are needed. 

\begin{acknowledgements} 

The work of J. M. and G. R. was partially supported by {\em Ministerio de Econom\'{\i}a y Competitividad} of Spain (project TEC2012-38883-C02-01 COMPREHENSION) and the Office of Naval Research Global (award no. N62909- 15-1-2011. D. C. and J. M. would also like to acknowledge the support of the Isaac Newton Institute through the program ``Monte Carlo Inference for High-Dimensional Statistical Models''. The authors would like to thank Dr. Katrin Achutegui for her valuable assistance in obtaining the numerical results displayed in Section \ref{sLorenz}.

\end{acknowledgements}

\appendix

\section{Proof of Lemma \ref{lmUnbiased}} \label{apUnbiased}

We proceed by induction in the time index $t$. For $t=0$, $\rho_0=\tau_0=\pi_0$ and, since $x_0^{(i)}$, $i = 1, ..., N$, are drawn from $\pi_0$, the equality $E[(f,\rho_0^N)] = (f,\rho_0)$ is straightforward.

Let us assume that 
\begin{equation}
E\left[
	(f,\rho_{t-1}^N)
\right] = (f,\rho_{t-1})
\label{eqInductionHypo}
\end{equation} 
for some $t>0$ and any $f \in B(\mX)$. If we use $\bar \mF_t$ to denote the $\sigma$-algebra generated by the set of random variables $\{ x_{0:t-1}^{(i)}, \bar x_{1:t}^{(i)} : 1 \le i \le N \}$ then we readily find that
\begin{equation}
E\left[
	(f,\rho_t^N) | \bar \mF_t 
\right] = E\left[
	G_t^N (f, \pi_t^N) | \bar \mF_t 
\right] = 
G_t^N (f, \bar \pi_t^N), \label{eqStep1}
\end{equation}
since $G_t^N$ is measurable w.r.t. $\bar \mF_t$ and $E[(f,\pi_t^N)|\bar \mF_t] = (f,\bar \pi_t^N)$. Moreover, if we recall that
\begin{equation}
(f,\bar \pi_t^N) = \sum_{i=1}^N w_t^{(i)} f(\bar x_t^{(i)}) = \sum_{i=1}^N \frac{
	g_t^{y_t}(\bar x_t^{(i)}) f(\bar x_t^{(i)})
}{
	\sum_{j=1}^N g_t^{y_t}(\bar x_t^{(j)})
}  = \frac{
	(fg_t^{y_t},\xi_t^N)
}{
	(g_t^{y_t},\xi_t^N)
}\nonumber
\end{equation}
then it is apparent from the definition of $G_t^N$ in \eqref{eqGtN} that
\begin{equation}
G_t^N (f, \bar \pi_t^N) = G_{t-1}^N (fg_t^{y_t},\xi_t^N).
\label{eqReduceG}
\end{equation}
Taking together \eqref{eqStep1} and \eqref{eqReduceG} we have
\begin{equation}
E\left[
	(f,\rho_t^N) | \bar \mF_t 
\right] = G_{t-1}^N (fg_t^{y_t},\xi_t^N).
\label{eqEndStep1}
\end{equation}

Let $\mF_{t-1}$ be the $\sigma$-algebra generated by the set of variables $\{ x_{0:t-1}^{(i)}, \bar x_{0:t-1}^{(i)} : 1 \le i \le N \}$. Since $\mF_{t-1} \subseteq \bar \mF_t$, Eq. \eqref{eqEndStep1} yields
\begin{eqnarray}
E\left[
	(f,\rho_t^N) | \mF_{t-1}
\right] &=& E\left[
	G_{t-1}^N (fg_t^{y_t},\xi_t^N) | \mF_{t-1}
\right] \nonumber \\
&=& G_{t-1}^N E\left[
	(fg_t^{y_t},\xi_t^N) | \mF_{t-1}
\right],
\label{eqBeginStep2}
\end{eqnarray}
since $G_{t-1}^N$ is measurable w.r.t. $\mF_{t-1}$. Moreover, for any $h \in B(\mX)$, it is straightforward to show that 
$$
E[ (h,\xi_t^N) | \mF_{t-1} ] = (h, \tau_t\pi_{t-1}^N) = \left(
	(h,\tau_t), \pi_{t-1}^N
\right),
$$
hence, as $f g_t^{y_t} \in B(\mX)$, we readily obtain
\begin{equation}
E\left[
	(fg_t^{y_t},\xi_t^N) | \mF_{t-1}
\right] = \left(
	(fg_t^{y_t},\tau_t), \pi_{t-1}^N
\right).
\label{eqBit1}
\end{equation}
Substituting \eqref{eqBit1} into \eqref{eqBeginStep2} we arrive at
\begin{eqnarray}
E\left[
	(f,\rho_t^N) | \mF_{t-1}
\right] &=& G_{t-1}^N  \left(
	(fg_t^{y_t},\tau_t), \pi_{t-1}^N
\right) \nonumber \\
&=& \left(
	(fg_t^{y_t},\tau_t), \rho_{t-1}^N
\right) \label{eqBit2},
\end{eqnarray}
where \eqref{eqBit2} follows from the definition of the estimate of $\rho_{t-1}$, namely $\rho_{t-1}^N = G_{t-1}^N\pi_{t-1}^N$. If we take unconditional expectations on both sides of Eq. \eqref{eqBit2}, we obtain 
\begin{eqnarray}
E\left[
	(f,\rho_t^N)
\right] &=& E\left[
	\left(
		(fg_t^{y_t},\tau_t), \rho_{t-1}^N
	\right)
\right] \nonumber \\
&=& 	\left(
	(fg_t^{y_t},\tau_t), \rho_{t-1}
\right) \label{eqBit3} \\
&=& (f, g_t^{y_t} \cdot \tau_t \rho_{t-1}) \label{eqBit4}\\
&=& (f,\rho_t), \label{eqEndStep2}
\end{eqnarray}
where equality \eqref{eqBit3} follows from the induction hypothesis \eqref{eqInductionHypo}, \eqref{eqBit4} is obtained by simply re-ordering \eqref{eqBit3} and Eq. \eqref{eqEndStep2} follows from the recursive definition of $\rho_t$ in \eqref{eqDefRhoRec}.

\section{Proof of Lemma \ref{lmLpRho}} \label{apLpRho}

For $t=0$, $\rho_0^N = \pi_0^N$, hence the result follows from Lemma \ref{lmConvBF}. At any time $t>0$, since $\rho_t^N = G_t^N \pi_t^N$, we readily have
\begin{eqnarray}
E\left[
	\left|
		(f,\rho_t^N) - (f,\rho_t)
	\right|^p
\right] &=& E\left[
	\left|
		\frac{1}{N} \sum_{i=1}^N G_t^N f(x_t^{(i)}) - (f,\rho_t)
	\right|^p
\right] \nonumber \\
&=& E\left[
	\left|
		\frac{1}{N} \sum_{i=1}^N Z_t^{(i)}
	\right|^p
\right], \label{eqToken}
\end{eqnarray}
where $Z_t^{(i)} =  G_t^N f(x_t^{(i)}) - (f,\rho_t)$, $i=1, ..., N$. It is apparent that the random variables $Z_t^{(i)}$, $i=1, ..., N$, are conditionally independent given the $\sigma$-algebra $\bar \mF_t$ generated by the set $\{ x_{0:t-1}^{(j)}, \bar x_{0:t}^{(j)} : 1 \le j \le N \}$. It can also be proved that every $Z_t^{(i)}$ is centred and bounded, as explicitly shown in the sequel. 

To see that $Z_t^{(i)}$ has zero mean, let us note first that 
\begin{equation}
E\left[
	G_t^N f(x_t^{(i)}) | \bar \mF_t 
\right] = G_t^N (f,\bar \pi_t^N),
\nonumber
\end{equation}
since $G_t^N$ is measurable w.r.t. $\bar \mF_t$. Moreover, by the same argument as in the proof of Lemma \ref{lmUnbiased}, one can show that $G_t^N(f,\bar \pi_t^N) = G_{t-1}^N (fg_t^{y_t},\xi_t^N)$ and, therefore,
\begin{eqnarray}
E\left[
	G_t^N f(x_t^{(i)}) | \mF_{t-1} 
\right] &=& E\left[
	G_{t-1}^N (fg_t^{y_t},\xi_t^N) | \mF_{t-1} 
\right] \nonumber \\
&=& G_{t-1}^N \left(
	(fg_t^{y_t},\tau_t), \pi_{t-1}^N
\right),
\label{eqToken0}
\end{eqnarray}
where we have used the fact that, for any $h \in B(\mX)$, $E[ (h,\xi_t^N) | \mF_{t-1} ] = ((h,\tau_t),\pi_{t-1}^N)$. However, since $\rho_{t-1}^N = G_{t-1}^N \pi_{t-1}^N$, Eq. \eqref{eqToken0} amounts to 
\begin{equation}
E\left[
	G_t^N f(x_t^{(i)}) | \mF_{t-1} 
\right] = \left(
	(fg_t^{y_t},\tau_t),\rho_{t-1}^N
\right)
\nonumber
\end{equation}
and taking (unconditional) expectations on both sides of the equation above yields
\begin{eqnarray}
E\left[
	G_t^N f(x_t^{(i)}) 
\right] &=& E\left[
	\left(
		(fg_t^{y_t},\tau_t),\rho_{t-1}^N
	\right)
\right] \nonumber\\
&=& \left(
	(fg_t^{y_t},\tau_t),\rho_{t-1}
\right) \label{eqToken1} \\
&=& (f,\rho_t), \label{eqToken2}
\end{eqnarray}
where \eqref{eqToken1} follows from Lemma \ref{lmUnbiased} (i.e., $\rho_{t-1}^N$ is unbiased) and \eqref{eqToken2} is a straightforward consequence of the definition of $\rho_t$ in \eqref{eqDefRhoRec}. Equation \eqref{eqToken2} states that $E[ Z_t^{(i)} ] = E[ G_t^N f(x_t^{(i)}) - (f,\rho_t) ] = 0$.

To see that (every) $Z_t^{(i)}$ is bounded, note that, for any finite $t$,
\begin{equation}
G_t^N \le \prod_{k=1}^t \| g_k^{y_k} \|_\infty < \infty,
\label{eqToken3}
\end{equation}
whereas 
\begin{eqnarray}
(f,\rho_t) &=& ((fg_t^{y_t},\tau_t),\rho_{t-1}) \nonumber\\
&=& ((((fg_t^{y_t},\tau_t)g_{t-1}^{y_{t-1}},\tau_{t-1})g_{t-2}^{y_{t-2}}, ..., \tau_1), \pi_0) \nonumber\\
&\le& \|f \|_\infty \prod_{k=1}^t \| g_k^{y_k} \|_\infty < \infty. \label{eqToken4}
\end{eqnarray}
Taking \eqref{eqToken3} and \eqref{eqToken4} together we arrive at
\begin{equation}
| Z_t^{(i)} | \le 2\| f \|_\infty \prod_{k=1}^t \| g_k^{y_k} \|_\infty
\label{eqToken5}
\end{equation}
which is finite for any finite $t$ (indeed, for every $t \le T$).

Since the variables $Z_t^{(i)}$, $i=1, ..., N$, in \eqref{eqToken} are bounded, with zero mean and conditionally independent given $\bar \mF_t$, it is not difficult to show (see, e.g., \cite[Lemma A.1]{Crisan13}) that  
\begin{equation}
E\left[
	\left|
		(f,\rho_t^N) - (f,\rho_t)
	\right|^p
\right] \le \frac{
	2^p \breve c_t^p \| f \|_\infty^p \prod_{k=1}^t \| g_k^{y_k} \|_\infty^p
}{
	N^\frac{p}{2}
},
\label{eqToken6}
\end{equation}
where the constant $\breve c_t$ is finite and independent of $N$. From \eqref{eqToken6} we easily obtain the inequality \eqref{eqStatementLpRho} in the statement of Lemma \ref{lmLpRho}, with $\tilde c_t =  2 \breve c_t \| f \|_\infty \prod_{k=1}^t \| g_k^{y_k} \|_\infty < \infty$ for any $t \le T < \infty$.

\bibliographystyle{plain}
\bibliography{bibliografia}




\end{document}